\documentclass{article}%
\usepackage{amsfonts}
\usepackage{amsmath}
\usepackage{amssymb}
\usepackage{graphicx}%
\makeatletter
\def\diagram{\m@th\leftwidth=\z@ \rightwidth=\z@ \topheight=\z@
\botheight=\z@ \setbox\@picbox\hbox\bgroup}
\def\enddiagram{\egroup\wd\@picbox\rightwidth\unitlength
\ht\@picbox\topheight\unitlength \dp\@picbox\botheight\unitlength
\hskip\leftwidth\unitlength\box\@picbox}
\def\bfig{\begin{diagram}}
\def\efig{\end{diagram}}
\newcount\wideness \newcount\leftwidth \newcount\rightwidth
\newcount\highness \newcount\topheight \newcount\botheight
\def\ratchet#1#2{\ifnum#1<#2 \global #1=#2 \fi}
\def\putbox(#1,#2)#3{\horsize{\wideness}{#3} \divide\wideness by 2
{\advance\wideness by #1 \ratchet{\rightwidth}{\wideness}}
{\advance\wideness by -#1 \ratchet{\leftwidth}{\wideness}}
\vertsize{\highness}{#3} \divide\highness by 2
{\advance\highness by #2 \ratchet{\topheight}{\highness}}
{\advance\highness by -#2 \ratchet{\botheight}{\highness}}
\put(#1,#2){\makebox(0,0){$#3$}}}
\def\putlbox(#1,#2)#3{\horsize{\wideness}{#3}
{\advance\wideness by #1 \ratchet{\rightwidth}{\wideness}}
{\ratchet{\leftwidth}{-#1}}
\vertsize{\highness}{#3} \divide\highness by 2
{\advance\highness by #2 \ratchet{\topheight}{\highness}}
{\advance\highness by -#2 \ratchet{\botheight}{\highness}}
\put(#1,#2){\makebox(0,0)[l]{$#3$}}}
\def\putrbox(#1,#2)#3{\horsize{\wideness}{#3}
{\ratchet{\rightwidth}{#1}}
{\advance\wideness by -#1 \ratchet{\leftwidth}{\wideness}}
\vertsize{\highness}{#3} \divide\highness by 2
{\advance\highness by #2 \ratchet{\topheight}{\highness}}
{\advance\highness by -#2 \ratchet{\botheight}{\highness}}
\put(#1,#2){\makebox(0,0)[r]{$#3$}}}
\def\adjust[#1]{}
\newcount \coefa
\newcount \coefb
\newcount \coefc
\newcount\tempcounta
\newcount\tempcountb
\newcount\tempcountc
\newcount\tempcountd
\newcount\xext
\newcount\yext
\newcount\xoff
\newcount\yoff
\newcount\gap
\newcount\arrowtypea
\newcount\arrowtypeb
\newcount\arrowtypec
\newcount\arrowtyped
\newcount\arrowtypee
\newcount\height
\newcount\width
\newcount\xpos
\newcount\ypos
\newcount\run
\newcount\rise
\newcount\arrowlength
\newcount\halflength
\newcount\arrowtype
\newdimen\tempdimen
\newdimen\xlen
\newdimen\ylen
\newsavebox{\tempboxa}
\newsavebox{\tempboxb}
\newsavebox{\tempboxc}
\newdimen\w@dth
\def\setw@dth#1#2{\setbox\z@\hbox{\m@th$#1$}\w@dth=\wd\z@
\setbox\@ne\hbox{\m@th$#2$}\ifnum\w@dth<\wd\@ne \w@dth=\wd\@ne \fi
\advance\w@dth by 1.2em}
\def\t@^#1_#2{\allowbreak\def\n@one{#1}\def\n@two{#2}\mathrel
{\setw@dth{#1}{#2}
\mathop{\hbox to \w@dth{\rightarrowfill}}\limits
\ifx\n@one\empty\else ^{\box\z@}\fi
\ifx\n@two\empty\else _{\box\@ne}\fi}}
\def\t@@^#1{\@ifnextchar_{\t@^{#1}}{\t@^{#1}_{}}}
\def\to{\@ifnextchar^{\t@@}{\t@@^{}}}
\def\t@left^#1_#2{\def\n@one{#1}\def\n@two{#2}\mathrel{\setw@dth{#1}{#2}
\mathop{\hbox to \w@dth{\leftarrowfill}}\limits
\ifx\n@one\empty\else ^{\box\z@}\fi
\ifx\n@two\empty\else _{\box\@ne}\fi}}
\def\t@@left^#1{\@ifnextchar_{\t@left^{#1}}{\t@left^{#1}_{}}}
\def\toleft{\@ifnextchar^{\t@@left}{\t@@left^{}}}
\def\two@^#1_#2{\allowbreak
\def\n@one{#1}\def\n@two{#2}\mathrel{\setw@dth{#1}{#2}
\mathop{\vcenter{\lineskip\z@\baselineskip\z@
\hbox to \w@dth{\rightarrowfill}                 \hbox to \w@dth{\rightarrowfill}}       }\limits
\ifx\n@one\empty\else ^{\box\z@}\fi
\ifx\n@two\empty\else _{\box\@ne}\fi}}
\def\tw@@^#1{\@ifnextchar _{\two@^{#1}}{\two@^{#1}_{}}}
\def\two{\@ifnextchar ^{\tw@@}{\tw@@^{}}}
\def\tofr@^#1_#2{\def\n@one{#1}\def\n@two{#2}\mathrel{\setw@dth{#1}{#2}
\mathop{\vcenter{\hbox to \w@dth{\rightarrowfill}\kern-1.7ex
\hbox to \w@dth{\leftarrowfill}}       }\limits
\ifx\n@one\empty\else ^{\box\z@}\fi
\ifx\n@two\empty\else _{\box\@ne}\fi}}
\def\t@fr@^#1{\@ifnextchar_ {\tofr@^{#1}}{\tofr@^{#1}_{}}}
\def\tofro{\@ifnextchar^ {\t@fr@}{\t@fr@^{}}}

\def\mon{\mathop{\m@th\hbox to
14.6\P@{\lasyb\char'51\hskip-2.1\P@$\arrext$\hss
$\mathord\rightarrow$}}\limits}
\def\leftmono{\mathrel{\m@th\hbox to
14.6\P@{$\mathord\leftarrow$\hss$\arrext$\hskip-2.1\P@\lasyb\char'50}}\limits}
\mathchardef\arrext="0200
\def\settypes(#1,#2,#3){\arrowtypea#1 \arrowtypeb#2 \arrowtypec#3}
\def\settoheight#1#2{\setbox\@tempboxa\hbox{#2}#1\ht\@tempboxa\relax}
\def\settodepth#1#2{\setbox\@tempboxa\hbox{#2}#1\dp\@tempboxa\relax}
\def\settokens`#1`#2`#3`#4`{     \def\tokena{#1}\def\tokenb{#2}\def\tokenc{#3}\def\tokend{#4}}
\def\setsqparms[#1`#2`#3`#4;#5`#6]{\arrowtypea #1
\arrowtypeb #2
\arrowtypec #3
\arrowtyped #4
\width #5
\height #6
}
\def\setpos(#1,#2){\xpos=#1 \ypos#2}
\def\settriparms[#1`#2`#3;#4]{\settripairparms[#1`#2`#3`1`1;#4]}
\def\settripairparms[#1`#2`#3`#4`#5;#6]{\arrowtypea #1
\arrowtypeb #2
\arrowtypec #3
\arrowtyped #4
\arrowtypee #5
\width #6
\height #6
}
\def\resetparms{\settripairparms[1`1`1`1`1;500]\width 500}
\resetparms
\def\mvector(#1,#2)#3{\put(0,0){\vector(#1,#2){#3}}\put(0,0){\vector(#1,#2){26}}}
\def\evector(#1,#2)#3{{\arrowlength #3
\put(0,0){\vector(#1,#2){\arrowlength}}\advance \arrowlength by-30
\put(0,0){\vector(#1,#2){\arrowlength}}}}
\def\horsize#1#2{\settowidth{\tempdimen}{$#2$}#1=\tempdimen
\divide #1 by\unitlength
}
\def\vertsize#1#2{\settoheight{\tempdimen}{$#2$}#1=\tempdimen
\settodepth{\tempdimen}{$#2$}\advance #1 by\tempdimen
\divide #1 by\unitlength
}
\def\putvector(#1,#2)(#3,#4)#5#6{{\ifnum3<\arrowtype
\putdashvector(#1,#2)(#3,#4)#5\arrowtype
\else
\ifnum\arrowtype<-3
\putdashvector(#1,#2)(#3,#4)#5\arrowtype
\else
\xpos=#1
\ypos=#2
\run=#3
\rise=#4
\arrowlength=#5
\ifnum \arrowtype<0
\ifnum \run=0
\advance \ypos by-\arrowlength
\else
\tempcounta \arrowlength
\multiply \tempcounta by\rise
\divide \tempcounta by\run
\ifnum\run>0
\advance \xpos by\arrowlength
\advance \ypos by\tempcounta
\else
\advance \xpos by-\arrowlength
\advance \ypos by-\tempcounta
\fi
\fi
\multiply \arrowtype by-1
\multiply \rise by-1
\multiply \run by-1
\fi
\ifcase \arrowtype
\or \put(\xpos,\ypos){\vector(\run,\rise){\arrowlength}}\or \put(\xpos,\ypos){\mvector(\run,\rise)\arrowlength}\or \put(\xpos,\ypos){\evector(\run,\rise){\arrowlength}}\fi\fi\fi
}}
\def\putsplitvector(#1,#2)#3#4{\xpos #1
\ypos #2
\arrowtype #4
\halflength #3
\arrowlength #3
\gap 140
\advance \halflength by-\gap
\divide \halflength by2
\ifnum\arrowtype>0
\ifcase \arrowtype
\or \put(\xpos,\ypos){\line(0,-1){\halflength}}       \advance\ypos by-\halflength
\advance\ypos by-\gap
\put(\xpos,\ypos){\vector(0,-1){\halflength}}   \or \put(\xpos,\ypos){\line(0,-1)\halflength}       \put(\xpos,\ypos){\vector(0,-1)3}       \advance\ypos by-\halflength
\advance\ypos by-\gap
\put(\xpos,\ypos){\vector(0,-1){\halflength}}   \or \put(\xpos,\ypos){\line(0,-1)\halflength}       \advance\ypos by-\halflength
\advance\ypos by-\gap
\put(\xpos,\ypos){\evector(0,-1){\halflength}}   \fi
\else \arrowtype=-\arrowtype
\ifcase\arrowtype
\or \advance \ypos by-\arrowlength
\put(\xpos,\ypos){\line(0,1){\halflength}}       \advance\ypos by\halflength
\advance\ypos by\gap
\put(\xpos,\ypos){\vector(0,1){\halflength}}   \or \advance \ypos by-\arrowlength
\put(\xpos,\ypos){\line(0,1)\halflength}       \put(\xpos,\ypos){\vector(0,1)3}       \advance\ypos by\halflength
\advance\ypos by\gap
\put(\xpos,\ypos){\vector(0,1){\halflength}}   \or \advance \ypos by-\arrowlength
\put(\xpos,\ypos){\line(0,1)\halflength}       \advance\ypos by\halflength
\advance\ypos by\gap
\put(\xpos,\ypos){\evector(0,1){\halflength}}   \fi
\fi
}
\def\putmorphism(#1)(#2,#3)[#4`#5`#6]#7#8#9{{\run #2
\rise #3
\ifnum\rise=0
\puthmorphism(#1)[#4`#5`#6]{#7}{#8}#9\else\ifnum\run=0
\putvmorphism(#1)[#4`#5`#6]{#7}{#8}#9\else
\setpos(#1)\arrowlength #7
\arrowtype #8
\ifnum\run=0
\else\ifnum\rise=0
\else
\ifnum\run>0
\coefa=1
\else
\coefa=-1
\fi
\ifnum\arrowtype>0
\coefb=0
\coefc=-1
\else
\coefb=\coefa
\coefc=1
\arrowtype=-\arrowtype
\fi
\width=2
\multiply \width by\run
\divide \width by\rise
\ifnum \width<0  \width=-\width\fi
\advance\width by60
\if l#9 \width=-\width\fi
\putbox(\xpos,\ypos){#4}{\multiply \coefa by\arrowlength\advance\xpos by\coefa
\multiply \coefa by\rise
\divide \coefa by\run
\advance \ypos by\coefa
\putbox(\xpos,\ypos){#5} }{\multiply \coefa by\arrowlength\divide \coefa by2
\advance \xpos by\coefa
\advance \xpos by\width
\multiply \coefa by\rise
\divide \coefa by\run
\advance \ypos by\coefa
\if l#9   \putrbox(\xpos,\ypos){#6}\else\if r#9   \putlbox(\xpos,\ypos){#6}\fi\fi }{\multiply \rise by-\coefc\multiply \run by-\coefc
\multiply \coefb by\arrowlength
\advance \xpos by\coefb
\multiply \coefb by\rise
\divide \coefb by\run
\advance \ypos by\coefb
\multiply \coefc by70
\advance \ypos by\coefc
\multiply \coefc by\run
\divide \coefc by\rise
\advance \xpos by\coefc
\multiply \coefa by140
\multiply \coefa by\run
\divide \coefa by\rise
\advance \arrowlength by\coefa
\ifcase\arrowtype
\or \put(\xpos,\ypos){\vector(\run,\rise){\arrowlength}}\or \put(\xpos,\ypos){\mvector(\run,\rise){\arrowlength}}\or \put(\xpos,\ypos){\evector(\run,\rise){\arrowlength}}\fi}\fi\fi\fi\fi}}
\newcount\numbdashes \newcount\lengthdash \newcount\increment
\def\howmanydashes{\numbdashes=\arrowlength \lengthdash=40
\divide\numbdashes by \lengthdash
\lengthdash=\arrowlength
\divide\lengthdash by \numbdashes
\increment=\lengthdash
\multiply\lengthdash by 3
\divide\lengthdash by 5
}
\def\putdashvector(#1)(#2,#3)#4#5{\ifnum#3=0 \putdashhvector(#1){#4}#5
\else
\ifnum#2=0
\putdashvvector(#1){#4}#5\fi\fi}
\def\putdashhvector(#1,#2)#3#4{{\arrowlength=#3 \howmanydashes
\multiput(#1,#2)(\increment,0){\numbdashes}{\vrule height .4pt width \lengthdash\unitlength}
\arrowtype=#4 \xpos=#1
\ifnum\arrowtype<0 \advance\arrowtype by 7 \fi
\ifcase\arrowtype
\or \advance\xpos by 10
\put(\xpos,#2){\vector(-1,0){\lengthdash}}
\advance\xpos by 40
\put(\xpos,#2){\vector(-1,0){\lengthdash}}
\or \advance \xpos by 10
\put(\xpos,#2){\vector(-1,0){\lengthdash}}
\advance\xpos by  \arrowlength
\advance\xpos by  -50
\put(\xpos,#2){\vector(-1,0){\lengthdash}}
\or \advance\xpos by 10
\put(\xpos,#2){\vector(-1,0){\lengthdash}}
\or \advance\xpos by \arrowlength
\advance\xpos by -\lengthdash
\put(\xpos,#2){\vector(1,0){\lengthdash}}
\or {\advance\xpos by 10
\put(\xpos,#2){\vector(1,0){\lengthdash}}}
\advance\xpos by \arrowlength
\advance\xpos by -\lengthdash
\put(\xpos,#2){\vector(1,0){\lengthdash}}
\or \advance\xpos by \arrowlength
\advance\xpos by -\lengthdash
\put(\xpos,#2){\vector(1,0){\lengthdash}}
\advance\xpos by -40
\put(\xpos,#2){\vector(1,0){\lengthdash}}
\fi
}}
\def\putdashvvector(#1,#2)#3#4{{\arrowlength=#3 \howmanydashes
\ypos=#2 \advance\ypos by -\arrowlength
\multiput(#1,#2)(0,\increment){\numbdashes}    {\vrule width .4pt height \lengthdash\unitlength}
\arrowtype=#4 \ypos=#2
\ifnum\arrowtype<0 \advance\arrowtype by 7 \fi
\ifcase\arrowtype
\or \advance\ypos by \arrowlength \advance\ypos by -40
\put(#1,\ypos){\vector(0,1){\lengthdash}}
\advance\ypos by -40
\put(#1,\ypos){\vector(0,1){\lengthdash}}
\or \advance\ypos by 10
\put(#1,\ypos){\vector(0,1){\lengthdash}}
\advance\ypos by \arrowlength \advance\ypos by -40
\put(#1,\ypos){\vector(0,1){\lengthdash}}
\or \advance\ypos by \arrowlength \advance\ypos by -40
\put(#1,\ypos){\vector(0,1){\lengthdash}}
\or \advance\ypos by 10
\put(#1,\ypos){\vector(0,-1){\lengthdash}}
\or \advance\ypos by 10
\put(#1,\ypos){\vector(0,-1){\lengthdash}}
\advance\ypos by \arrowlength \advance\ypos by -40
\put(#1,\ypos){\vector(0,-1){\lengthdash}}
\or \advance\ypos by 10
\put(#1,\ypos){\vector(0,-1){\lengthdash}}
\advance\ypos by 40
\put(#1,\ypos){\vector(0,-1){\lengthdash}}
\fi
}}
\def\puthmorphism(#1,#2)[#3`#4`#5]#6#7#8{{\xpos #1
\ypos #2
\width #6
\arrowlength #6
\arrowtype=#7
\putbox(\xpos,\ypos){#3\vphantom{#4}}{\advance \xpos by\arrowlength
\putbox(\xpos,\ypos){\vphantom{#3}#4}}\horsize{\tempcounta}{#3}\horsize{\tempcountb}{#4}\divide \tempcounta by2
\divide \tempcountb by2
\advance \tempcounta by30
\advance \tempcountb by30
\advance \xpos by\tempcounta
\advance \arrowlength by-\tempcounta
\advance \arrowlength by-\tempcountb
\putvector(\xpos,\ypos)(1,0)\arrowlength\arrowtype
\divide \arrowlength by2
\advance \xpos by\arrowlength
\vertsize{\tempcounta}{#5}\divide\tempcounta by2
\advance \tempcounta by20
\if a#8    \advance \ypos by\tempcounta
\putbox(\xpos,\ypos){#5}\else
\advance \ypos by-\tempcounta
\putbox(\xpos,\ypos){#5}\fi}}
\def\putvmorphism(#1,#2)[#3`#4`#5]#6#7#8{{\xpos #1
\ypos #2
\arrowlength #6
\arrowtype #7
\settowidth{\xlen}{$#5$}\putbox(\xpos,\ypos){#3}{\advance \ypos by-\arrowlength
\putbox(\xpos,\ypos){#4}}{\advance\arrowlength by-140
\advance \ypos by-70
\ifdim\xlen>0pt
\if m#8      \putsplitvector(\xpos,\ypos)\arrowlength\arrowtype
\else
\putvector(\xpos,\ypos)(0,-1)\arrowlength\arrowtype
\fi
\else
\putvector(\xpos,\ypos)(0,-1)\arrowlength\arrowtype
\fi}\ifdim\xlen>0pt
\divide \arrowlength by2
\advance\ypos by-\arrowlength
\if l#8      \advance \xpos by-40
\putrbox(\xpos,\ypos){#5}   \else\if r#8      \advance \xpos by40
\putlbox(\xpos,\ypos){#5}   \else
\putbox(\xpos,\ypos){#5}   \fi\fi
\fi
}}
\def\putsquarep<#1>(#2)[#3;#4`#5`#6`#7]{{\setsqparms[#1]\setpos(#2)\settokens`#3`\puthmorphism(\xpos,\ypos)[\tokenc`\tokend`{#7}]{\width}{\arrowtyped}b\advance\ypos by \height
\puthmorphism(\xpos,\ypos)[\tokena`\tokenb`{#4}]{\width}{\arrowtypea}a\putvmorphism(\xpos,\ypos)[``{#5}]{\height}{\arrowtypeb}l\advance\xpos by \width
\putvmorphism(\xpos,\ypos)[``{#6}]{\height}{\arrowtypec}r}}
\def\putsquare{\@ifnextchar <{\putsquarep}{\putsquarep   <\arrowtypea`\arrowtypeb`\arrowtypec`\arrowtyped;\width`\height>}}
\def\square{\@ifnextchar< {\squarep}{\squarep
<\arrowtypea`\arrowtypeb`\arrowtypec`\arrowtyped;\width`\height>}}
\def\squarep<#1>[#2`#3`#4`#5;#6`#7`#8`#9]{{\setsqparms[#1]\diagram\putsquarep<\arrowtypea`\arrowtypeb`\arrowtypec`\arrowtyped;\width`\height>(0,0)[#2`#3`#4`{#5};#6`#7`#8`{#9}]\enddiagram}}
\def\putptrianglep<#1>(#2,#3)[#4`#5`#6;#7`#8`#9]{{\settriparms[#1]\xpos=#2 \ypos=#3
\advance\ypos by \height
\puthmorphism(\xpos,\ypos)[#4`#5`{#7}]{\height}{\arrowtypea}a\putvmorphism(\xpos,\ypos)[`#6`{#8}]{\height}{\arrowtypeb}l\advance\xpos by\height
\putmorphism(\xpos,\ypos)(-1,-1)[``{#9}]{\height}{\arrowtypec}r}}
\def\putptriangle{\@ifnextchar <{\putptrianglep}{\putptrianglep
<\arrowtypea`\arrowtypeb`\arrowtypec;\height>}}
\def\ptriangle{\@ifnextchar <{\ptrianglep}{\ptrianglep
<\arrowtypea`\arrowtypeb`\arrowtypec;\height>}}
\def\ptrianglep<#1>[#2`#3`#4;#5`#6`#7]{{\settriparms[#1]\diagram\putptrianglep<\arrowtypea`\arrowtypeb`\arrowtypec;\height>(0,0)[#2`#3`#4;#5`#6`{#7}]\enddiagram}}
\def\putqtrianglep<#1>(#2,#3)[#4`#5`#6;#7`#8`#9]{{\settriparms[#1]\xpos=#2 \ypos=#3
\advance\ypos by\height
\puthmorphism(\xpos,\ypos)[#4`#5`{#7}]{\height}{\arrowtypea}a\putmorphism(\xpos,\ypos)(1,-1)[``{#8}]{\height}{\arrowtypeb}l\advance\xpos by\height
\putvmorphism(\xpos,\ypos)[`#6`{#9}]{\height}{\arrowtypec}r}}
\def\putqtriangle{\@ifnextchar <{\putqtrianglep}{\putqtrianglep
<\arrowtypea`\arrowtypeb`\arrowtypec;\height>}}
\def\qtriangle{\@ifnextchar <{\qtrianglep}{\qtrianglep
<\arrowtypea`\arrowtypeb`\arrowtypec;\height>}}
\def\qtrianglep<#1>[#2`#3`#4;#5`#6`#7]{{\settriparms[#1]\width=\height                                \diagram\putqtrianglep<\arrowtypea`\arrowtypeb`\arrowtypec;\height>(0,0)[#2`#3`#4;#5`#6`{#7}]\enddiagram}}
\def\putdtrianglep<#1>(#2,#3)[#4`#5`#6;#7`#8`#9]{{\settriparms[#1]\xpos=#2 \ypos=#3
\puthmorphism(\xpos,\ypos)[#5`#6`{#9}]{\height}{\arrowtypec}b\advance\xpos by \height \advance\ypos by\height
\putmorphism(\xpos,\ypos)(-1,-1)[``{#7}]{\height}{\arrowtypea}l\putvmorphism(\xpos,\ypos)[#4``{#8}]{\height}{\arrowtypeb}r}}
\def\putdtriangle{\@ifnextchar <{\putdtrianglep}{\putdtrianglep
<\arrowtypea`\arrowtypeb`\arrowtypec;\height>}}
\def\dtriangle{\@ifnextchar <{\dtrianglep}{\dtrianglep
<\arrowtypea`\arrowtypeb`\arrowtypec;\height>}}
\def\dtrianglep<#1>[#2`#3`#4;#5`#6`#7]{{\settriparms[#1]\width=\height                                \diagram\putdtrianglep<\arrowtypea`\arrowtypeb`\arrowtypec;\height>(0,0)[#2`#3`#4;#5`#6`{#7}]\enddiagram}}
\def\putbtrianglep<#1>(#2,#3)[#4`#5`#6;#7`#8`#9]{{\settriparms[#1]\xpos=#2 \ypos=#3
\puthmorphism(\xpos,\ypos)[#5`#6`{#9}]{\height}{\arrowtypec}b\advance\ypos by\height
\putmorphism(\xpos,\ypos)(1,-1)[``{#8}]{\height}{\arrowtypeb}r\putvmorphism(\xpos,\ypos)[#4``{#7}]{\height}{\arrowtypea}l}}
\def\putbtriangle{\@ifnextchar <{\putbtrianglep}{\putbtrianglep
<\arrowtypea`\arrowtypeb`\arrowtypec;\height>}}
\def\btriangle{\@ifnextchar <{\btrianglep}{\btrianglep
<\arrowtypea`\arrowtypeb`\arrowtypec;\height>}}
\def\btrianglep<#1>[#2`#3`#4;#5`#6`#7]{{\settriparms[#1]\width=\height                               \diagram\putbtrianglep<\arrowtypea`\arrowtypeb`\arrowtypec;\height>(0,0)[#2`#3`#4;#5`#6`{#7}]\enddiagram}}
\def\putAtrianglep<#1>(#2,#3)[#4`#5`#6;#7`#8`#9]{{\settriparms[#1]\xpos=#2 \ypos=#3
{\multiply \height by2
\puthmorphism(\xpos,\ypos)[#5`#6`{#9}]{\height}{\arrowtypec}b}\advance\xpos by\height \advance\ypos by\height
\putmorphism(\xpos,\ypos)(-1,-1)[#4``{#7}]{\height}{\arrowtypea}l\putmorphism(\xpos,\ypos)(1,-1)[``{#8}]{\height}{\arrowtypeb}r}}
\def\putAtriangle{\@ifnextchar <{\putAtrianglep}{\putAtrianglep
<\arrowtypea`\arrowtypeb`\arrowtypec;\height>}}
\def\Atriangle{\@ifnextchar <{\Atrianglep}{\Atrianglep
<\arrowtypea`\arrowtypeb`\arrowtypec;\height>}}
\def\Atrianglep<#1>[#2`#3`#4;#5`#6`#7]{{\settriparms[#1]\width=\height                                     \diagram\putAtrianglep<\arrowtypea`\arrowtypeb`\arrowtypec;\height>(0,0)[#2`#3`#4;#5`#6`{#7}]\enddiagram}}
\def\putAtrianglepairp<#1>(#2)[#3;#4`#5`#6`#7`#8]{{\settripairparms[#1]\setpos(#2)\settokens`#3`\puthmorphism(\xpos,\ypos)[\tokenb`\tokenc`{#7}]{\height}{\arrowtyped}b\advance\xpos by\height
\puthmorphism(\xpos,\ypos)[\phantom{\tokenc}`\tokend`{#8}]{\height}{\arrowtypee}b\advance\ypos by\height
\putmorphism(\xpos,\ypos)(-1,-1)[\tokena``{#4}]{\height}{\arrowtypea}l\putvmorphism(\xpos,\ypos)[``{#5}]{\height}{\arrowtypeb}m\putmorphism(\xpos,\ypos)(1,-1)[``{#6}]{\height}{\arrowtypec}r}}
\def\putAtrianglepair{\@ifnextchar <{\putAtrianglepairp}{\putAtrianglepairp   <\arrowtypea`\arrowtypeb`\arrowtypec`\arrowtyped`\arrowtypee;\height>}}
\def\Atrianglepair{\@ifnextchar <{\Atrianglepairp}{\Atrianglepairp   <\arrowtypea`\arrowtypeb`\arrowtypec`\arrowtyped`\arrowtypee;\height>}}
\def\Atrianglepairp<#1>[#2;#3`#4`#5`#6`#7]{{\settripairparms[#1]\settokens`#2`\width=\height                                \diagram\putAtrianglepairp                            <\arrowtypea`\arrowtypeb`\arrowtypec`\arrowtyped`\arrowtypee;\height>(0,0)[{#2};#3`#4`#5`#6`{#7}]\enddiagram}}
\def\putVtrianglep<#1>(#2,#3)[#4`#5`#6;#7`#8`#9]{{\settriparms[#1]\xpos=#2 \ypos=#3
\advance\ypos by\height
{\multiply\height by2
\puthmorphism(\xpos,\ypos)[#4`#5`{#7}]{\height}{\arrowtypea}a}\putmorphism(\xpos,\ypos)(1,-1)[`#6`{#8}]{\height}{\arrowtypeb}l\advance\xpos by\height
\advance\xpos by\height
\putmorphism(\xpos,\ypos)(-1,-1)[``{#9}]{\height}{\arrowtypec}r}}
\def\putVtriangle{\@ifnextchar <{\putVtrianglep}{\putVtrianglep
<\arrowtypea`\arrowtypeb`\arrowtypec;\height>}}
\def\Vtriangle{\@ifnextchar <{\Vtrianglep}{\Vtrianglep
<\arrowtypea`\arrowtypeb`\arrowtypec;\height>}}
\def\Vtrianglep<#1>[#2`#3`#4;#5`#6`#7]{{\settriparms[#1]\width=\height                                 \diagram\putVtrianglep<\arrowtypea`\arrowtypeb`\arrowtypec;\height>(0,0)[#2`#3`#4;#5`#6`{#7}]\enddiagram}}
\def\putVtrianglepairp<#1>(#2)[#3;#4`#5`#6`#7`#8]{{
\settripairparms[#1]\setpos(#2)\settokens`#3`\advance\ypos by\height
\putmorphism(\xpos,\ypos)(1,-1)[`\tokend`{#6}]{\height}{\arrowtypec}l\puthmorphism(\xpos,\ypos)[\tokena`\tokenb`{#4}]{\height}{\arrowtypea}a\advance\xpos by\height
\puthmorphism(\xpos,\ypos)[\phantom{\tokenb}`\tokenc`{#5}]{\height}{\arrowtypeb}a\putvmorphism(\xpos,\ypos)[``{#7}]{\height}{\arrowtyped}m\advance\xpos by\height
\putmorphism(\xpos,\ypos)(-1,-1)[``{#8}]{\height}{\arrowtypee}r}}
\def\putVtrianglepair{\@ifnextchar <{\putVtrianglepairp}{\putVtrianglepairp    <\arrowtypea`\arrowtypeb`\arrowtypec`\arrowtyped`\arrowtypee;\height>}}
\def\Vtrianglepair{\@ifnextchar <{\Vtrianglepairp}{\Vtrianglepairp    <\arrowtypea`\arrowtypeb`\arrowtypec`\arrowtyped`\arrowtypee;\height>}}
\def\Vtrianglepairp<#1>[#2;#3`#4`#5`#6`#7]{{\settripairparms[#1]\settokens`#2`\diagram\putVtrianglepairp                             <\arrowtypea`\arrowtypeb`\arrowtypec`\arrowtyped`\arrowtypee;\height>(0,0)[{#2};#3`#4`#5`#6`{#7}]\enddiagram}}
\def\putCtrianglep<#1>(#2,#3)[#4`#5`#6;#7`#8`#9]{{\settriparms[#1]\xpos=#2 \ypos=#3
\advance\ypos by\height
\putmorphism(\xpos,\ypos)(1,-1)[``{#9}]{\height}{\arrowtypec}l\advance\xpos by\height
\advance\ypos by\height
\putmorphism(\xpos,\ypos)(-1,-1)[#4`#5`{#7}]{\height}{\arrowtypea}l{\multiply\height by 2
\putvmorphism(\xpos,\ypos)[`#6`{#8}]{\height}{\arrowtypeb}r}}}
\def\putCtriangle{\@ifnextchar <{\putCtrianglep}{\putCtrianglep
<\arrowtypea`\arrowtypeb`\arrowtypec;\height>}}
\def\Ctriangle{\@ifnextchar <{\Ctrianglep}{\Ctrianglep
<\arrowtypea`\arrowtypeb`\arrowtypec;\height>}}
\def\Ctrianglep<#1>[#2`#3`#4;#5`#6`#7]{{\settriparms[#1]\width=\height                               \diagram\putCtrianglep<\arrowtypea`\arrowtypeb`\arrowtypec;\height>(0,0)[#2`#3`#4;#5`#6`{#7}]\enddiagram}}
\def\putDtrianglep<#1>(#2,#3)[#4`#5`#6;#7`#8`#9]{{\settriparms[#1]\xpos=#2 \ypos=#3
\advance\xpos by\height \advance\ypos by\height
\putmorphism(\xpos,\ypos)(-1,-1)[``{#9}]{\height}{\arrowtypec}r\advance\xpos by-\height \advance\ypos by\height
\putmorphism(\xpos,\ypos)(1,-1)[`#5`{#8}]{\height}{\arrowtypeb}r{\multiply\height by 2
\putvmorphism(\xpos,\ypos)[#4`#6`{#7}]{\height}{\arrowtypea}l}}}
\def\putDtriangle{\@ifnextchar <{\putDtrianglep}{\putDtrianglep
<\arrowtypea`\arrowtypeb`\arrowtypec;\height>}}
\def\Dtriangle{\@ifnextchar <{\Dtrianglep}{\Dtrianglep
<\arrowtypea`\arrowtypeb`\arrowtypec;\height>}}
\def\Dtrianglep<#1>[#2`#3`#4;#5`#6`#7]{{\settriparms[#1]\width=\height                              \diagram\putDtrianglep<\arrowtypea`\arrowtypeb`\arrowtypec;\height>(0,0)[#2`#3`#4;#5`#6`{#7}]\enddiagram}}
\def\setrecparms[#1`#2]{\width=#1 \height=#2}
\def\recursep<#1`#2>[#3;#4`#5`#6`#7`#8]{{\m@th
\width=#1 \height=#2
\settokens`#3`
\settowidth{\tempdimen}{$\tokena$}
\ifdim\tempdimen=0pt
\savebox{\tempboxa}{\hbox{$\tokenb$}}  \savebox{\tempboxb}{\hbox{$\tokend$}}  \savebox{\tempboxc}{\hbox{$#6$}}\else
\savebox{\tempboxa}{\hbox{$\hbox{$\tokena$}\times\hbox{$\tokenb$}$}}  \savebox{\tempboxb}{\hbox{$\hbox{$\tokena$}\times\hbox{$\tokend$}$}}  \savebox{\tempboxc}{\hbox{$\hbox{$\tokena$}\times\hbox{$#6$}$}}\fi
\ypos=\height
\divide\ypos by 2
\xpos=\ypos
\advance\xpos by \width
\bfig
\putCtrianglep<-1`1`1;\ypos>(0,0)[`\tokenc`;#5`#6`{#7}]\puthmorphism(\ypos,0)[\tokend`\usebox{\tempboxb}`{#8}]{\width}{-1}b\puthmorphism(\ypos,\height)[\tokenb`\usebox{\tempboxa}`{#4}]{\width}{-1}a\advance\ypos by \width
\putvmorphism(\ypos,\height)[``\usebox{\tempboxc}]{\height}1r\efig
}}
\def\recurse{\@ifnextchar <{\recursep}{\recursep<\width`\height>}}
\def\puttwohmorphisms(#1,#2)[#3`#4;#5`#6]#7#8#9{{\puthmorphism(#1,#2)[#3`#4`]{#7}0a
\ypos=#2
\advance\ypos by 20
\puthmorphism(#1,\ypos)[\phantom{#3}`\phantom{#4}`#5]{#7}{#8}a
\advance\ypos by -40
\puthmorphism(#1,\ypos)[\phantom{#3}`\phantom{#4}`#6]{#7}{#9}b
}}
\def\puttwovmorphisms(#1,#2)[#3`#4;#5`#6]#7#8#9{{\putvmorphism(#1,#2)[#3`#4`]{#7}0a
\xpos=#1
\advance\xpos by -20
\putvmorphism(\xpos,#2)[\phantom{#3}`\phantom{#4}`#5]{#7}{#8}l
\advance\xpos by 40
\putvmorphism(\xpos,#2)[\phantom{#3}`\phantom{#4}`#6]{#7}{#9}r
}}
\def\puthcoequalizer(#1)[#2`#3`#4;#5`#6`#7]#8#9{{\setpos(#1)\puttwohmorphisms(\xpos,\ypos)[#2`#3;#5`#6]{#8}11\advance\xpos by #8
\puthmorphism(\xpos,\ypos)[\phantom{#3}`#4`#7]{#8}1{#9}
}}
\def\putvcoequalizer(#1)[#2`#3`#4;#5`#6`#7]#8#9{{\setpos(#1)\puttwovmorphisms(\xpos,\ypos)[#2`#3;#5`#6]{#8}11\advance\ypos by -#8
\putvmorphism(\xpos,\ypos)[\phantom{#3}`#4`#7]{#8}1{#9}
}}
\def\putthreehmorphisms(#1)[#2`#3;#4`#5`#6]#7(#8)#9{{\setpos(#1) \settypes(#8)
\if a#9      \vertsize{\tempcounta}{#5}     \vertsize{\tempcountb}{#6}     \ifnum \tempcounta<\tempcountb \tempcounta=\tempcountb \fi
\else
\vertsize{\tempcounta}{#4}     \vertsize{\tempcountb}{#5}     \ifnum \tempcounta<\tempcountb \tempcounta=\tempcountb \fi
\fi
\advance \tempcounta by 60
\puthmorphism(\xpos,\ypos)[#2`#3`#5]{#7}{\arrowtypeb}{#9}
\advance\ypos by \tempcounta
\puthmorphism(\xpos,\ypos)[\phantom{#2}`\phantom{#3}`#4]{#7}{\arrowtypea}{#9}
\advance\ypos by -\tempcounta \advance\ypos by -\tempcounta
\puthmorphism(\xpos,\ypos)[\phantom{#2}`\phantom{#3}`#6]{#7}{\arrowtypec}{#9}
}}
\def\setarrowtoks[#1`#2`#3`#4`#5`#6]{\def\toka{#1}
\def\tokb{#2}
\def\tokc{#3}
\def\tokd{#4}
\def\toke{#5}
\def\tokf{#6}
}
\def\hex{\@ifnextchar <{\hexp}{\hexp<1000`400>}}
\def\hexp<#1`#2>[#3`#4`#5`#6`#7`#8;#9]{\setarrowtoks[#9]
\yext=#2 \advance \yext by #2
\xext=#1 \advance\xext by \yext
\bfig
\putCtriangle<-1`0`1;#2>(0,0)[`#5`;\tokb``\tokd]
\xext=#1 \yext=#2 \advance \yext by #2
\putsquare<1`0`0`1;\xext`\yext>(#2,0)[#3`#4`#7`#8;\toka```\tokf]
\advance \xext by #2
\putDtriangle<0`1`-1;#2>(\xext,0)[`#6`;`\tokc`\toke]
\efig
}
\makeatother
\newtheorem{theorem}{Theorem}

\newtheorem{corollary}[theorem]{Corollary}

\newtheorem{definition}[theorem]{Definition}

\newtheorem{lemma}[theorem]{Lemma}

\newtheorem{proposition}[theorem]{Proposition}

\newenvironment{proof}[1][Proof]{\textbf{#1.} }{\ \rule{0.5em}{0.5em}}
\begin{document}

\title{Kochen-Specker theorem for\\von Neumann algebras}
\author{Andreas D\"{o}ring\\Fachbereich Mathematik\\Johann Wolfgang Goethe-Universit\"{a}t \\Robert-Mayer-Stra\ss e 10\\60054 Frankfurt, Germany\\mail: adoering@math.uni-frankfurt.de}
\maketitle

\begin{abstract}
The Kochen-Specker theorem has been discussed intensely ever since its
original proof in 1967. It is one of the central no-go theorems of quantum
theory, showing the non-existence of a certain kind of hidden states models.
In this paper, we first offer a new, non-combinatorial proof for quantum
systems with a type $I_{n}$ factor as algebra of observables, including
$I_{\infty}$. Afterwards, we give a proof of the Kochen-Specker theorem for an
arbitrary von Neumann algebra $\mathcal{R}$ without summands of types $I_{1}$
and $I_{2}$, using a known result on two-valued measures on the projection
lattice $\mathcal{P(R)}$. Some connections with presheaf formulations as
proposed by Isham and Butterfield are made.

\end{abstract}

\newpage

\section{Introduction}

In quantum theory, including quantum mechanics in the von Neumann
representation, quantum field theory and quantum information theory,
observables are represented by self-adjoint operators $A$ in some von Neumann
algebra $\mathcal{R}$, the \emph{algebra of observables}. The algebra
$\mathcal{R}$ is contained in $\mathcal{L(H)}$, the set of bounded linear
operators on some separable Hilbert space $\mathcal{H}$. The self-adjoint
operators $\mathcal{R}_{sa}$ form a real linear space in the algebra
$\mathcal{R}$. It is important for the interpretation of quantum theory to see
if there is a possibility to assign a \emph{value} to each observable such
that (i)\ an observable $A\in\mathcal{R}_{sa}$ is assigned one element of its
spectrum and (ii) if for two observables $A,B\in\mathcal{R}_{sa}$ one has
$B=g(A)$ for some (Borel) function $g$, then the value assigned to $B$, say
$b$, is given as $g(a)$, where $a$ is the value assigned to $A$. If this were
possible, one could imagine to build some realist model of the quantum world
where all observables have definite values, like in classical
mechanics.\newline

The first condition, namely that each observable should be assigned one of its
spectral values, is quite obvious. The second condition implements the fact
that the observables are not all independent. In fact, for every abelian von
Neumann algebra $\mathcal{M}$ (think of some abelian subalgebra of
$\mathcal{R}$), there is a self-adjoint operator $A$ generating $\mathcal{M}$,
i.e. $\mathcal{M}=\{A,I\}^{\prime\prime}$. This was already proved by von
Neumann (\cite{vNeu32}). For a modern reference, see Thm. III.1.21 in
\cite{TakI79}. Every operator $B\in\mathcal{M}$ is a Borel function of $A$,
$B=g(A)$. One has $g(\operatorname*{sp}A)\subseteq\operatorname*{sp}B$ (see
Ch. 5.2 in \cite{KadRinI97}). If $g$ is continuous or if $\mathcal{H}$ is
finite-dimensional, equality holds. It is natural to demand that the spectral
value $b$ assigned to $B$ is given as $g(a)$, where $a$ is the value assigned
to $A$. This condition is often called the \emph{FUNC principle}.\newline

Kochen and Specker started from a related question in their classical paper
\cite{KocSpe67}: is there a space of \emph{hidden states}? A hidden state
$\psi$ would be given by a probability measure $\mu_{\psi}$ on a generalized
quantum mechanical phase space $\Omega$, such that an observable $A$ is given
as a mapping%
\[
f_{A}:\Omega\longrightarrow\mathbb{R},
\]
a \emph{hidden variable}. When the system is in the hidden state $\psi$ and
the observable $A$ is measured, the probability to find a value $r$ lying in
the Borel set $U$ is required to be%
\begin{equation}
P_{A,\psi}(U)=\mu_{\psi}(f_{A}^{-1}(U)). \tag{1}%
\end{equation}
Furthermore, the expectation value $E_{\psi}(A)$ of $A$ when the system is in
the hidden state $\psi$ is required to be%
\[
E_{\psi}(A)=\int_{\Omega}f_{a}(\omega)d\mu_{\psi}(\omega).
\]
Kochen and Specker demonstrate that it is trivial to construct such a
generalized phase space $\Omega$ if functional relations between the
observables are neglected, but the problem really starts when one takes these
relations into account. If $B=g(A)$ for some observables $A,B\in
\mathcal{R}_{sa}$ and a Borel function $g$, one should have%
\begin{equation}
f_{B}=f_{g(A)}=g\circ f_{A}. \tag{2}%
\end{equation}
This simply translates the functional relation between the operators $A$ and
$B$ into the corresponding relation between the hidden variables $f_{A}$ and
$f_{B}$. Since $B=g(A)$ can only be if $A$ and $B$ commute and since every
abelian von Neumann algebra is generated by a self-adjoint operator, Kochen
and Specker go on to introduce \emph{partial algebras}, where algebraic
relations are defined exclusively between observables that are
\emph{commeasurable}. If one regards a von Neumann algebra $\mathcal{R}$ as
the algebra of observables, as we do here, $\mathcal{R}$ is a partial algebra
in an obvious way: one just keeps the algebraic relations between commuting
operators and neglects the algebraic relations between non-commuting
operators, since at first sight (2) is a condition on commuting operators only
and those are commeasurable. This point seems important, because a hidden
variable no-go theorem by J. von Neumann (\cite{vNeu32}) has been criticized
(see e.g. \cite{Bell66}) for the fact that von Neumann required additivity to
be preserved even between non-commuting operators, which does not seem
adequate in the light of (2). However, in fact (2) does \emph{not} just pose
conditions on commuting, but also on non-commuting operators. The reason is
that typically an observable $B$ is given as a function $B=g(A)=h(C)$ of
non-commuting observables $A,C\in\mathcal{R}_{sa}$. We will see that in this
way (2) becomes a very strong condition, ruling out hidden states models of
the kind described above.\newline

An abelian subalgebra $\mathcal{M}$ of $\mathcal{R}$ is often called a
\emph{context} in the physics literature. The self-adjoint operators
$\mathcal{M}_{sa}$ in a maximal context $\mathcal{M}$ form a maximal set of
commeasurable observables. Condition (2) seems to be a condition within each
context solely, but in fact it is a condition \textquotedblright
across\ contexts\textquotedblright, because each observable $B$ typically is
contained in many contexts.\newline

The elements of the hypothetical generalized phase space $\Omega$ would be
generalized pure states. In a slight abuse of language, $\Omega$ is also
called the space of hidden states. If one assumes that there is some space
$\Omega$ of hidden states, such that (1) and (2) are satisfied and that there
is an embedding $f:\mathcal{R}_{sa}\rightarrow\mathbb{R}^{\Omega}$ of the
quantum mechanical observables into the mappings from $\Omega$ to $\mathbb{R}%
$, one would have a lot of valuations as described above, assigning a spectral
value to each observable and preserving functional relations: every point
$\omega\in\Omega$ defines such a valuation $v$ by%
\[
v(A):=f_{A}(\omega).
\]
Demonstrating that there are no such valuations (Kochen and Specker called
them \emph{prediction functions}) thus shows that there is no space $\Omega$
of hidden states as described above. More directly, the non-existence of
valuation functions means that no realist interpretation of quantum mechanics
is possible which assumes that all the observables have definite values at the
same time.\newline

It is a funny fact that in spite of many references to it, there seems to be
no single result called \emph{the} Kochen-Specker theorem. Above, we tried to
lay out (very roughly, admittedly) the train of thought in \cite{KocSpe67},
and it seems sensible to spell out the Kochen-Specker theorem as
follows:\newline

\textbf{Kochen-Specker theorem}: Let $\mathcal{R}$ $\simeq\mathcal{L(H)}%
,\ \dim\mathcal{H}\geq3$ be the algebra of observables of some quantum system
($\mathcal{R}$ is a type $I_{n}$ factor, $n=\dim\mathcal{H}$). There is no
space $\Omega$ of hidden states such that (1) and (2) are satisfied, i.e.
there is no realist phase space model of quantum theory assigning spectral
values to all observables at once, preserving functional relations between
them.\newline

It is not obvious at first sight if the Kochen-Specker theorem holds for more
general von Neumann algebras $\mathcal{R}$, since each $\mathcal{R}$ that is
not a type $I_{n}$ factor is properly contained in some $\mathcal{L(H)}$, and
so there are less conditions (encoded in the FUNC principle) than for
$\mathcal{L(H)}$ itself. This might lead to speculation if some hidden states,
realist model of quantum systems with an observable algebra $\mathcal{R}$
other than a type $I_{n}$ factor exists. A necessary condition would be the
existence of a valuation function $v:\mathcal{R}_{sa}\rightarrow\mathbb{R}%
$.\newline

To prove the non-existence of valuation functions, Kochen and Specker
(\cite{KocSpe67}) concentrate on the projections $\mathcal{P(R)}$ of
$\mathcal{R}$. These form a partial Boolean algebra. A valuation function $v$
can assign $0$ or $1$ to a projection $E$, since $\operatorname*{sp}%
E=\{0,1\}$. Kochen and Specker examine if the partial Boolean algebra
$\mathcal{P(R)}$ can be embedded into a Boolean algebra. The existence of such
an embedding is a necessary condition for the existence of a valuation
function. For the case $\mathcal{H}=\mathbb{R}^{3},\mathcal{R}=\mathbb{M}%
_{n}(3)$, they construct a finitely generated subalgebra $D\subset\mathcal{R}$
that cannot be embedded into a Boolean algebra, thus showing that there is no
valuation function in this case. Kochen and Specker use 117 vectors in their
construction, corresponding to 117 projections onto one-dimensional subspaces.
Later on, this number could be reduced to 33 by A. Peres and 31 by Conway and
Kochen, see \cite{Per93} and references therein. The proofs are combinatorial
in nature, giving a counterexample.\newline

In this paper, we will use another approach. Let $\mathcal{R}$ be a von
Neumann algebra. Assuming the existence of a valuation function $v:\mathcal{R}%
_{sa}\rightarrow\mathbb{R}$ (see Def. \ref{DValFunc} below), we show that $v$
induces a so-called \emph{quasi-state} $v^{\prime}:\mathcal{R}\rightarrow
\mathbb{C}$ (see Def. \ref{DQuasiState}) such that $v^{\prime}|_{\mathcal{R}%
_{sa}}=v$. This quasi-state is a pure state of every abelian subalgebra
$\mathcal{M}$ of $\mathcal{R}$. Restricting $v^{\prime}$ to the projections
$\mathcal{P(R)}$, we obtain a finitely additive probability measure on
$\mathcal{P(R)}$.\newline

If $\mathcal{R}$ is a type $I_{n}$ factor ($n\in\{3,4,...\}$), Gleason's
theorem (\cite{Gle57}) shows that $v^{\prime}$ is a state of $\mathcal{R}$ of
the form $v^{\prime}(\_)=tr(\rho\_)$. But such a state does \emph{not} assign
$0$ or $1$ to every projection, hence we have a contradiction of one of the
defining conditions of the valuation function $v$. The case of a type
$I_{\infty}$ factor can be treated easily.\newline

For more general von Neumann algebras, another proof is presented. Using the
Gleason-Christensen-Yeadon theorem (Thm. \ref{TGenGleas}), a generalization of
Gleason's theorem, we again show that $v^{\prime}$ is a state if $\mathcal{R}$
has no summand of type $I_{2}$. Hamhalter showed in \cite{Ham93} that every
finitely additive two-valued probability measure $\mu:\mathcal{P(R)}%
\rightarrow\{0,1\}$ gives rise to a \emph{multiplicative} state of
$\mathcal{R}$. Since $v^{\prime}|_{\mathcal{P(R)}}$ is of this kind, a
valuation function $v$ induces a multiplicative state $v^{\prime}$. If a von
Neumann algebra $\mathcal{R}$ contains no summand of type $I_{1}$, then there
are no multiplicative states of $\mathcal{R}$, so there is no valuation
function for a von Neumann algebra $\mathcal{R}$ without summands of types
$I_{1}$ and $I_{2}$. Thus, the generalized Kochen-Specker theorem holds for
all von Neumann algebras $\mathcal{R}$ without summands of types $I_{1}$ and
$I_{2}$.\newline

In section 3, we give two different reformulations of the generalized
Kochen-Specker theorem in the language of presheafs. For $\mathcal{R}%
=\mathcal{L(H)}$, this has been proposed by Isham, Butterfield and Hamilton
(\cite{IshBut98,IshBut99,HIB00,IshBut02}). They observed that the FUNC
principle means that certain presheafs on small categories have global
sections.\newline

Our first presheaf formulation of the Kochen-Specker theorem, generalizing the
category and presheaf chosen in \cite{HIB00}, is closely related to our first
proof (subsections 2.1, 2.2). The second formulation uses another presheaf
which does have global sections: each state of the von$\ $Neumann algebra
$\mathcal{R}$ induces one. However, the Kochen-Specker theorem means that
there are no global sections of the kind a valuation function would induce,
giving a pure state of every abelian subalgebra $\mathcal{M}$ of $\mathcal{R}$.

\section{The new proofs}

\subsection{Valuation functions and quasi-states}

All von Neumann and $C^{\ast}$-algebras treated here are unital subalgebras of
some $\mathcal{L(H)}$, the algebra of bounded operators on a Hilbert space
$\mathcal{H}$. $\mathcal{R}_{sa}$ denotes the real linear space of
self-adjoint elements of a $C^{\ast}$- or von Neumann algebra $\mathcal{R}$,
$\mathcal{P(R)}$ is the lattice of projections of $\mathcal{R}$.

\begin{definition}
\label{DProbMeasures}Let $\mathcal{R\subseteq L(H)}$ be a von Neumann algebra.
A \textbf{finitely additive probability measure} $\mu$ is a mapping from
$\mathcal{P(R)}$ to $\mathbb{R}$ such that%
\begin{align*}
&  \text{(M1) }\forall E\in\mathcal{P(R)}:0\leq\mu(E)\leq1\text{ and }%
\mu(I)=1,\\
&  \text{(M2) If }E,F\in\mathcal{P(R)}\text{ such that }EF=0\text{, then }%
\mu(E\vee F)=\mu(E)+\mu(F)\text{.}%
\end{align*}
If in addition to (M2) one of the stronger conditions%
\begin{align*}
\text{(M2}\sigma\text{) }  &  \mu(\bigvee_{n\in I}P_{n})=\sum_{n\in I}%
\mu(P_{n})\text{ for every countable family }\\
&  \{P_{n}\}_{n\in I}\text{ of orthogonal projections in }\mathcal{P(R)}%
\text{,}%
\end{align*}%
\begin{align*}
\text{(M2c) }  &  \mu(\bigvee_{j\in J}P_{j})=\sum_{j\in J}\mu(P_{j})\text{ for
every family }\{P_{j}\}_{j\in J}\text{ }\\
&  \text{of orthogonal projections in }\mathcal{P(R)}\text{ }%
\end{align*}
holds, then $\mu$ is called a $\mathbf{\sigma}$\textbf{-additive (countably
additive) }or a \textbf{completely additive probability measure}, respectively.
\end{definition}

Every \emph{normal} state $\phi:\mathcal{R}\rightarrow\mathbb{C}$ is of the
form $\phi(\_)=tr(\rho\_)$ for some positive trace class operator of trace
$1$, see Thm. 7.1.12 in \cite{KadRinII97}. Such a normal state induces a
completely additive probability measure by restriction to $\mathcal{P(R)}$.
For type $I$ factors, the converse is also true, as Gleason showed in his
classical paper \cite{Gle57}. For ease of reference, we cite Gleason's theorem:

\begin{theorem}
\label{TGleason}(Gleason 1957) Let $\mathcal{R}$ be a type $I_{n}$ factor,
$n\in\{3,4,...\},$ $\mathcal{R\simeq L(H)},$\newline$\dim\mathcal{H}=n$, and
let $\mu$ be a finitely additive probability measure on $\mathcal{P(R)}$.
There is some positive trace class operator $\rho$ of trace $1$ such that%
\begin{equation}
\forall E\in\mathcal{P(R)}:\mu(E)=tr(\rho E). \tag{1}%
\end{equation}

If $\mathcal{H}$ is infinite-dimensional and separable, $\mu$ is $\sigma
$-additive and $\mathcal{R}$ is isomorphic to the type $I_{\infty}$ factor
$\mathcal{L(H)}$, then there is some positive trace class operator $\rho$ of
trace $1$ such that (1) holds.

If $\mathcal{H}$ is an arbitrary infinite-dimensional Hilbert space (possibly
non-separable), $\mu$ is completely additive and $\mathcal{R}$ is isomorphic
to the type $I_{\infty}$ factor $\mathcal{L(H)}$, then there is some positive
trace class operator $\rho$ of trace $1$ such that (1) holds. (For this
partial result, see Thm. 2.3 in \cite{Mae89}.)
\end{theorem}

This classifies the probability measures on the projection lattices of type
$I$ factors. In particular, they all come from normal states of the form
$tr(\rho\_)$.\newline

From now on, we will assume that $\mathcal{H}$ is separable.\newline

We now give the precise definition of a valuation function, which is the
starting point for the proof of the Kochen-Specker theorem.

\begin{definition}
\label{DValFunc}Let $\mathcal{H}$ be a Hilbert space, $\mathcal{R}%
\subseteq\mathcal{L(H)}$ a von Neumann algebra. A \textbf{valuation function}
is a mapping $v:\mathcal{R}_{sa}\rightarrow\mathbb{R}$ such that\newline(a)
$v(A)\in\operatorname*{sp}A$ and\newline(b) for all Borel functions
$f:\mathbb{R\rightarrow R}$, one has $v(f(A))=f(v(A))$.
\end{definition}

Kochen and Specker call this a \emph{prediction function}, see \cite{KocSpe67}%
. $v(I)=1$ and $v(0)=0$ follow. Condition (a) is often called the
\textbf{Spectrum rule}, condition (b) is the \textbf{FUNC principle}.

\begin{definition}
\label{DValFuncExt}Let $v:\mathcal{R}_{sa}\rightarrow\mathbb{R}$ be a
valuation function. We extend $v$ in a canonical manner to a function%
\begin{align*}
v^{\prime}:\mathcal{R}  &  \longrightarrow\mathbb{C},\\
B=A_{1}+iA_{2}  &  \longmapsto v(A_{1})+iv(A_{2}),
\end{align*}
where $B=A_{1}+iA_{2}$ is the unique decomposition of $B$ into self-adjoint
operators $A_{1},A_{2}\in\mathcal{R}_{sa}$.
\end{definition}

Obviously, $v^{\prime}(A)=v(A)$ for a self-adjoint operator $A\in
\mathcal{R}_{sa}$. This will be used throughout.

\begin{lemma}
\label{LValFuncExtBorel}Let $g:\mathbb{R\rightarrow C}$ be a Borel function,
$g_{r}:\mathbb{R\rightarrow R}$ its real part, $g_{i}:\mathbb{R\rightarrow R}
$ its imaginary part, $g=g_{r}+ig_{i}$. Thus $g$ acts on $a\in\mathbb{R}$ as%
\[
g(a)=g_{r}(a)+ig_{i}(a)
\]
and on self-adjoint operators $A$ as%
\[
g(A)=g_{r}(A)+ig_{i}(A).
\]
Let $v:\mathcal{R}_{sa}\rightarrow\mathbb{R}$ be a valuation function,
$v^{\prime}:\mathcal{R}\rightarrow\mathbb{C}$ its extension. Then $v^{\prime
}(g(A))=g(v^{\prime}(A))$ holds for all self-adjoint operators $A\in
\mathcal{R}_{sa}$.
\end{lemma}

\begin{proof}
One has%
\begin{align*}
v^{\prime}(g(A))  &  =v^{\prime}(g_{r}(A)+ig_{i}(A))\\
&  =v(g_{r}(A))+iv(g_{i}(A))\\
&  =g_{r}(v(A))+ig_{i}(v(A))\\
&  =g(v(A))\\
&  =g(v^{\prime}(A)).
\end{align*}

\end{proof}

\begin{lemma}
\label{LValFuncExtChar}If $v:\mathcal{R}_{sa}\rightarrow\mathbb{R}$ is a
valuation function and $\mathcal{M}\subseteq\mathcal{R}$ is an abelian von
Neumann subalgebra, then $v^{\prime}|_{\mathcal{M}}$ \ is a character of
$\mathcal{M}$. $v|_{\mathcal{M}_{sa}}$ is a real-valued, $\mathbb{R}$-linear,
linear functional .
\end{lemma}

\begin{proof}
Let $A\in\mathcal{M}_{sa}$ be a self-adjoint operator that generates
$\mathcal{M}$, i.e. $\mathcal{M}=\{A,I\}^{\prime\prime}$ (see \cite[Prop.
III.1.21, p. 112]{TakI79}). All operators $B,C\in\mathcal{M}$ are Borel
functions of $A$:%
\[
B=f(A),\quad C=g(A),
\]
where $f,g:\mathbb{R\rightarrow C}$ are Borel functions on $\operatorname*{sp}%
A\subseteq\mathbb{R}$. Since $B+C\in\mathcal{M}$, there also is a Borel
function $h:\mathbb{R\rightarrow C}$ such that $B+C=f(A)+g(A)=:h(A)$ and hence%
\begin{align*}
v^{\prime}(B+C)  &  =v^{\prime}(f(A)+g(A))\\
&  =v^{\prime}(h(A))\\
&  =h(v^{\prime}(A))\\
&  =f(v^{\prime}(A))+g(v^{\prime}(A))\\
&  =v^{\prime}(f(A))+v^{\prime}(g(A))\\
&  =v^{\prime}(B)+v^{\prime}(C).
\end{align*}
Analogously for $BC=CB$: there is a Borel function $j:\mathbb{R\rightarrow C}$
such that $BC=f(A)g(A)=:j(A)$ and hence
\begin{align*}
v^{\prime}(BC)  &  =v^{\prime}(f(A)g(A))\\
&  =v^{\prime}(j(A))\\
&  =j(v^{\prime}(A))\\
&  =f(v^{\prime}(A))g(v^{\prime}(A))\\
&  =v^{\prime}(f(A))v^{\prime}(g(A))\\
&  =v^{\prime}(B)v^{\prime}(C).
\end{align*}
The $\mathbb{C}$-linearity of $v^{\prime}|_{\mathcal{M}}$ is obvious. Let
$\alpha\in\mathbb{C}$. Then%
\begin{align*}
v^{\prime}(\alpha B)  &  =v^{\prime}(\alpha f(A))\\
&  =v^{\prime}(k(A))\\
&  =k(v^{\prime}(A))\\
&  =\alpha f(v^{\prime}(A))\\
&  =\alpha v^{\prime}(f(A))\\
&  =\alpha v^{\prime}(B),
\end{align*}
where $k:=M_{\alpha}\circ f$. This shows that $v^{\prime}|_{\mathcal{M}}$ is a
character of $\mathcal{M}$. Restricting to $\mathcal{M}_{sa}$, one obtains a
real-valued, $\mathbb{R}$-linear, linear functional.
\end{proof}

\ 

A character (multiplicative linear functional)\ of an abelian $C^{\ast}%
$-algebra $\mathcal{M}$ is a pure state of $\mathcal{M}$. So the above lemma
shows that a valuation function induces a pure state on every abelian
subalgebra $\mathcal{M}\subseteq\mathcal{R}$, which is exactly what one would
expect from a physical point of view. Lemma \ref{LValFuncExtChar} is closely
related to the \emph{sum rule} and the \emph{product rule} first described in
\cite{Fin74}, see also \cite{Red87}.\newline

J. F. Aarnes has introduced the notion of a quasi-state on a $C^{\ast}%
$-algebra in his paper \cite{Aar69}:

\begin{definition}
\label{DQuasiState}Let $\mathcal{A}$ be a unital $C^{\ast}$-algebra. A
\textbf{quasi-state} of $\mathcal{A}$ is a functional $\rho$ satisfying the
following three conditions:

(1) For each $B\in\mathcal{A}_{sa}$, $\rho$ is linear and positive on the
abelian $C^{\ast}$-subalgebra $\mathcal{A}_{B}\subseteq\mathcal{A}$ generated
by $B$ and $I$.

(2) If $C=A_{1}+iA_{2}$ for self-adjoint $A_{1},A_{2}\in\mathcal{A}_{sa}$,
then $\rho(C)=\rho(A_{1})+i\rho(A_{2})$.

(3) $\rho(I)=1$.
\end{definition}

\begin{lemma}
\label{LValFuncExtQS}$v^{\prime}$ is a quasi-state.
\end{lemma}

\begin{proof}
(1) $v^{\prime}$ is linear on every abelian subalgebra $\mathcal{M}%
\subseteq\mathcal{R}$. $v^{\prime}$ is positive on each such $\mathcal{M}$
(and on the whole of $\mathcal{R}$), since a positive operator $B^{\ast}B$ is
assigned some element of its spectrum,$\ v^{\prime}(B^{\ast}B)=v(B^{\ast}B)\in
sp(B^{\ast}B)$.\newline

(2) For $B=A_{1}+iA_{2}$ $(A_{1},A_{2}\in\mathcal{R}_{sa})$ one has%
\[
v^{\prime}(B)=v(A_{1})+iv(A_{2})=v^{\prime}(A_{1})+iv^{\prime}(A_{2}),
\]
the former because of the definition of $v^{\prime}$, the latter because
$v^{\prime}(A)=v(A)$ for $A\in\mathcal{R}_{sa}$.\newline

(3) $v^{\prime}(I)=1$ holds.\newline

Thus $v^{\prime}$ is a quasi-state.
\end{proof}

\ 

A quasi-state of an abelian von Neumann algebra is a state. This follows from
the fact that an abelian von Neumann algebra on a separable Hilbert space is
generated by an operator $A$ (\cite[Prop. III.1.21]{TakI79}), a result we
already used in Lemma \ref{LValFuncExtChar}.\newline

It is easy to see that a quasi-state on an arbitrary von Neumann algebra
$\mathcal{R}$, when restricted to the lattice of projections $\mathcal{P(R)}$,
gives a finitely additive probability measure. We follow the proof given in
\cite[Cor. 7.9, p. 264]{Mae89}:

\begin{lemma}
\label{LQSProbM}If $\rho$ is a quasi-state of \ a von Neumann algebra
$\mathcal{R}$, then $\rho|_{\mathcal{P(R)}}$ is a finitely additive
probability measure $\mathcal{P(R)}\rightarrow\lbrack0,1]$.
\end{lemma}

\begin{proof}
For $E\in\mathcal{P(R)}$, we have $0=\rho(0)\leq\rho(E)\leq\rho(I)=1$, since
$\rho$ is positive on $\{E,I\}^{\prime\prime}$. If $EF=0$ for $E,F\in
\mathcal{P(R)}$, then $\mathcal{M}:=\{E,F,I\}^{\prime\prime}\subseteq
\mathcal{R}$ is abelian and $\rho|_{\mathcal{M}}$ is a state, in particular,
it is additive. Hence,%
\[
\rho(E\vee F)=\rho(E+F)=\rho(E)+\rho(F).
\]

\end{proof}

\ 

To clarify the relation between normal states and valuation functions, we will
need the following fact (Lemma 6.5.6 in \cite{KadRinII97}):

\begin{lemma}
\label{LPartingProjs}Let $\mathcal{R}$ be a von Neumann algebra with no
central portion of type $I$ (equivalently, with no non-zero abelian
projections), and let $E\in\mathcal{P(R)}$. For each positive integer $n$,
there are $n$ equivalent orthogonal projections with sum $E$.
\end{lemma}

\begin{lemma}
\label{LNormalStateNoValFunc}Let $\mathcal{R}\subseteq\mathcal{L(H)}$ be a von
Neumann algebra of type $I_{n},\ n\in\{2,3,...\}\cup\{\infty\}$, or of type
$II$ or $III$, and let $\phi$ be a normal state of $\mathcal{R}$. There is a
projection $E\in\mathcal{P(R)}$ such that $\phi(E)\notin
\{0,1\}=\operatorname*{sp}E$.
\end{lemma}

\begin{proof}
Since $\phi$ is a normal state, it is weakly continuous and of the form%
\[
\phi(\_)=tr(\rho\_)
\]
for some positive trace class operator $\rho$ of trace $1$, see \cite[Thm.
7.1.12]{KadRinII97}.\newline

We assume that $\phi(E)=tr(\rho E)\in\{0,1\}$ holds for all $E\in
\mathcal{P(R)}$. Let $\{e_{k}\}_{k\in K}$ be an orthonormal basis of
$\mathcal{H}$ that is adapted to $E$, i.e. for all $k$, $e_{k}\in
\operatorname*{im}E\cup\operatorname*{im}(I-E)$. Let $K_{E}:=\{k\in
K\ |\ e_{k}\in\operatorname*{im}E\}$. If $tr(\rho E)=1$, we have%
\begin{align*}
1  &  =\sum_{k}\left\langle \rho e_{k},Ee_{k}\right\rangle \\
&  =\sum_{k\in K_{E}}\left\langle \rho e_{k},e_{k}\right\rangle .
\end{align*}
Since $\left\langle \rho e_{k},e_{k}\right\rangle \geq0$ for all $k\in K$ and
$tr\rho=1$, we see that $\left\langle \rho e_{k},e_{k}\right\rangle =0$ for
all $k\in K\backslash K_{E}$, and hence $\rho e_{k}=0$ for all $k\in
K\backslash K_{E}$. Therefore, $\rho(I-E)=0$, i.e. $\rho=\rho E$ and%
\[
\rho E=\rho=\rho^{\ast}=E\rho.
\]
If $tr(\rho E)=0$, we have $tr(\rho(I-E))=1$, since $tr(\rho I)=1=tr(\rho
E)+tr(\rho(I-E))$. It follows that $\rho(I-E)=(I-E)\rho$ and thus $\rho
E=E\rho$ in this case, too. Since a von Neumann algebra is generated by its
projections, we obtain%
\[
(\forall E\in\mathcal{P(R)}:tr(\rho E)\in\{0,1\})\Longrightarrow\rho
\in\mathcal{R}^{\prime},
\]
where $\mathcal{R}^{\prime}$ is the commutant of $\mathcal{R}$. Now let
$\theta\in\mathcal{R}$ be a partial isometry such that $\theta^{\ast}\theta=E$
and $F:=\theta\theta^{\ast}$. One has%
\[
tr(\rho E)=tr(\rho\theta^{\ast}\theta)=tr(\theta\rho\theta^{\ast}%
)=tr(\theta\theta^{\ast}\rho)=tr(F\rho)=tr(\rho F),
\]
so from $E\sim F$ it follows that $\phi(E)=tr(\rho E)=tr(\rho F)=\phi
(F)$.\newline

If $\mathcal{R}$ is of type $I_{n}$, $n\geq2$, then the identity $I$ is the
sum of $n$ equivalent abelian (orthogonal) projections $E_{j}$ $(j=1,...,n)$.
We have%
\[
1=\phi(I)=\phi(\sum_{j=1}^{n}E_{j})=\sum_{j=1}^{n}\phi(E_{j}),
\]
so $\phi(E_{j})=\frac{1}{n}$ $(j=1,...,n)$, which contradicts our assumption
$\phi(E)\in\{0,1\}$ for all $E\in\mathcal{P(R)}$.\newline

If $\mathcal{R}$ is of type $I_{\infty}$, we use the halving lemma (6.3.3 in
\cite{KadRinII97}) to show that there is a projection $F\in\mathcal{P(R)}$
such that $F\sim F^{\perp}:=I-F$. If $\mathcal{R}$ is of type $II$ or $III$,
then we employ Lemma \ref{LPartingProjs} for $E=I$ and $n=2$ to obtain the
same. We have
\[
1=\phi(I)=\phi(F+I-F)=\phi(F)+\phi(F^{\perp}),
\]
so $\phi(F)=\phi(F^{\perp})=\frac{1}{2}$, which contradicts our assumption.
\end{proof}

\ 

This lemma means that restricting a normal state $\phi$ of a von Neumann
algebra $\mathcal{R}$ that is not of type $I_{1}$ (i.e. abelian) to
$\mathcal{R}_{sa}$ can never give a valuation function. The proof of this
lemma is based on the proof of Thm. 6.4 in \cite{deG01} (Thm.
\ref{TPureStateBooleanQP} in our paper).

\subsection{Type $I_{n}$ factors}

In this subsection, let $\mathcal{R}$ be a type $I$ factor.

\begin{definition}
\label{DBoolSecStoneSpec}Using some notions from \cite{deG01}, we call a
maximal distributive sublattice of $\mathcal{P(R)}$ a \textbf{Boolean sector
}$\mathbb{B}$. The abelian von Neumann algebra $\mathcal{M}(\mathbb{B)}%
\subseteq\mathcal{R}$ generated by $\mathbb{B}$ has Gelfand spectrum
$\Omega(\mathcal{M}(\mathbb{B))}$. One can define the \textbf{Stone spectrum}
$\mathcal{Q(}\mathbb{B)}$ of $\mathbb{B}$ as the set of maximal dual ideals in
$\mathbb{B}$, equipped with the topology induced by the sets%
\[
\mathcal{Q}_{E}(\mathbb{B)}:=\{\beta\in\mathcal{Q(}\mathbb{B)}\ |\ E\in
\beta\},
\]
see Ch. 4 in \cite{deG01} (the Stone spectrum of $\mathbb{B}$ is called the
Stone space of $\mathbb{B}$ there). The elements $\beta$ of the Stone spectrum
$\mathcal{Q}(\mathbb{B)}$ are called \textbf{quasipoints}. To every quasipoint
$\beta$ of $\mathcal{Q(}\mathbb{B)}$, there corresponds an element $\omega$ of
$\Omega(\mathcal{M(}\mathbb{B))}$, since the Stone spectrum and the Gelfand
spectrum are homeomorphic (\cite[Thm. 5.2]{deG01}). More generally, the Stone
spectrum $\mathcal{Q(M)}:=\mathcal{Q(P(M))}$ of an abelian von Neumann algebra
$\mathcal{M}$ is defined as the set of maximal dual ideals $\beta$ in
$\mathcal{P(M)}$, equipped with the topology induced by the sets%
\[
\mathcal{Q}_{E}(\mathcal{M}):=\{\beta\in\mathcal{Q(M)}\ |\ E\in\beta\}.
\]
$\mathcal{Q(M)}$ is homeomorphic to the Gelfand spectrum $\Omega(\mathcal{M})$.
\end{definition}

Every state $\rho$ of $\mathcal{R}$ induces a bounded positive Radon measure
$\mu_{\rho}^{\mathbb{B}}$ of norm $1$ on $\mathcal{Q(}\mathbb{B)}$ by%
\begin{align*}
\mu_{\rho}^{\mathbb{B}}:C(\mathcal{Q(}\mathbb{B))}  &  \mathbb{\longrightarrow
}\mathbb{C}\\
A  &  \longmapsto tr(\rho A).
\end{align*}

We will use the following result (Thm. 6.4 in \cite{deG01}):

\begin{theorem}
\label{TPureStateBooleanQP}Let $\rho$ be a state of $\mathcal{R}$, and let
$\mathbb{B}\subseteq\mathcal{P(R)}$ be a Boolean sector. Then the Radon
measure $\mu_{\rho}^{\mathbb{B}}$ on the Stone spectrum $\mathcal{Q(}%
\mathbb{B)}$ is the point measure $\varepsilon_{\beta_{0}}$ for some
$\beta_{0}\in\mathcal{Q(}\mathbb{B)}$, if and only if there is an $x\in
S^{1}(\mathcal{H})$ such that $\mathbb{C}x\in\mathbb{B},\ \beta_{0}%
=\beta_{\mathbb{C}x}$ and $\rho=P_{\mathbb{C}x}$. Here $\beta_{\mathbb{C}x}$
is the unique quasipoint containing $P_{\mathbb{C}x}$.
\end{theorem}

\begin{proposition}
\label{PNoValFunc}Let $\mathcal{R}$ be a factor of type $I_{n},\ n\in
\{3,4,...\}\cup\{\infty\}$. There is no valuation function $v:\mathcal{R}%
_{sa}\rightarrow\mathbb{R}$.
\end{proposition}

\begin{proof}
We first treat the case of finite $n$. As shown above, assuming the existence
of a valuation function $v$, one has a quasi-state $v^{\prime}$ of
$\mathcal{R}$, which is a pure state (i.e. a character) of every abelian von
Neumann subalgebra $\mathcal{M}\subseteq\mathcal{R}$. Lemma \ref{LQSProbM}
shows that $v^{\prime}|_{\mathcal{P(R)}}$ is a finitely additive probability
measure. Gleason's theorem (Thm. \ref{TGleason}) shows that $v^{\prime
}|_{\mathcal{P(R)}}$ comes from a state of $\mathcal{R\simeq L(H}%
_{n}\mathcal{)}$, where $\mathcal{H}_{n}$ is an $n$-dimensional Hilbert
space.\newline

The state given by Gleason's theorem is of the form $tr(\rho\_)$, where $\rho$
is a positive trace class operator of trace $1$. According to the spectral
theorem, every operator $A\in\mathcal{R}$ is the norm limit of complex linear
combinations of projections, hence there is a unique possibility to extend the
probability measure $v^{\prime}|_{\mathcal{P(R)}}$ to a state (given by
linearly extending $v^{\prime}|_{\mathcal{P(R)}}$). Of course, this extension
simply is $v^{\prime}$, so%
\[
v^{\prime}(\_)=tr(\rho\_).
\]
Let $\mathbb{B}$ be a \emph{Boolean sector} of $\mathcal{P(R)}$, and let
$\mathcal{M}(\mathbb{B)}$ be the maximal abelian subalgebra of $\mathcal{R}$
generated by $\mathbb{B}$. Since $v^{\prime}|_{\mathcal{M}(\mathbb{B)}}$ is a
pure state, it corresponds to exactly one element $\beta_{0}\in\mathcal{Q}%
(\mathbb{B)}$. $v^{\prime}$ induces a point measure on $\mathcal{M}%
(\mathbb{B)=}C\mathbb{(}\mathcal{Q}(\mathbb{B))}$ in this way.\newline

According to Thm. \ref{TPureStateBooleanQP}, $\rho$ and $\beta$ are of the
form%
\begin{align*}
\rho &  =P_{\mathbb{C}x},\\
\beta &  =\beta_{\mathbb{C}x,}%
\end{align*}
and thus%
\[
v^{\prime}(\_)=tr(P_{\mathbb{C}x}\_).
\]
This form does not depend on the chosen Boolean sector $\mathbb{B}$.\newline

Now let $\mathbb{B}^{\prime}$ be a different Boolean sector that does not
contain the projections $P_{\mathbb{C}x},I-P_{\mathbb{C}x}$. There is a
projection $F^{\prime}\in\mathbb{B}^{\prime}$ such that%
\[
v^{\prime}(F^{\prime})=tr(P_{\mathbb{C}x}F^{\prime})\notin\{0,1\},
\]
contradicting the defining condition $v^{\prime}(E)=v(E)\in\operatorname*{sp}%
E=\{0,1\}\ (E\in\mathcal{P(R)})$ of a valuation function. This shows that
there is no valuation function $v:\mathcal{R}_{sa}\rightarrow\mathbb{R}$ for
factors $\mathcal{R}$ of type $I_{n},\ n\in\{3,4,...\}$, from which the
Kochen-Specker theorem follows. Instead of referring to \cite{deG01}, we could
have used Lemma \ref{LNormalStateNoValFunc}.\newline

Now let $\mathcal{R}$ be a type $I_{\infty}$ factor, $\mathcal{R\simeq L(H)}$
for an infinite-dimensional separable Hilbert space $\mathcal{H}$.
$\mathcal{R}$ contains a subfactor of type $I_{n}$ for every $n\in
\{3,4,...\}$: Let $\mathcal{S}$ be a type $I_{n}$ factor, $\mathcal{S\simeq
L(H}_{n})$. The separable Hilbert spaces $\mathcal{H}_{n}\otimes\mathcal{H}$
and $\mathcal{H}$ are isomorphic and will be identified. Embed $\mathcal{L(H}%
_{n})$ into $\mathcal{L(H)}$ via the mapping%
\begin{align*}
\mathcal{L(H}_{n})  &  \longrightarrow\mathcal{L(H}_{n}\otimes\mathcal{H}%
)\simeq\mathcal{L(H)}\\
A  &  \longmapsto A\otimes I.
\end{align*}
This guarantees that the identity $I_{n}$ of $\mathcal{L(H}_{n})$ is mapped to
the identity $I$ of $\mathcal{L(H)}$.\newline

We assume that there is a valuation function $v:\mathcal{R}_{sa}%
\rightarrow\mathbb{R}$. Restricting $v$ to the self-adjoint part of a type
$I_{n}$ subfactor $\mathcal{S}$ of $\mathcal{R}$ gives a valuation function
for $\mathcal{S}$. Since we saw that there is no such valuation function,
there can be none for $\mathcal{R}_{sa}$.
\end{proof}

\subsection{Von Neumann algebras without type $I_{2}$ summand}

The proof for the type $I_{n}$ case proceeded in two steps: first, assuming
that there is a valuation function $v:\mathcal{R}_{sa}\rightarrow\mathbb{R}$
for $\mathcal{R}$ a type $I_{n}$ algebra, we showed that it induces a
quasi-state $v^{\prime}:\mathcal{R}\rightarrow\mathbb{C}$ in a canonical
manner and thus a finitely additive probability measure $v^{\prime
}|_{\mathcal{P(R)}}$. In a second step, we used Gleason's theorem to see that
$v^{\prime}$ is a \emph{state} of $\mathcal{R}$ of the form $v^{\prime
}(\_)=tr(\rho\_)$, which cannot satisfy the defining conditions for a
valuation function, because there are projections $F^{\prime}\in
\mathcal{P(R)}$ such that $v^{\prime}(F^{\prime})\notin\operatorname*{sp}%
F^{\prime}=\{0,1\}$.\newline

If we want to treat more general von Neumann algebras $\mathcal{R}$, we first
must assure that the quasi-state $v^{\prime}$ is a state of $\mathcal{R}$.
Since we know that $v^{\prime}|_{\mathcal{P(R)}}$ is a finitely additive
probability measure for an arbitrary von Neumann algebra $\mathcal{R}$ (Lemma
\ref{LQSProbM}), a generalization of Gleason's theorem is needed, showing that
this probability measure comes from a state. There is a beautiful and detailed
paper by S. Maeda \cite{Mae89} on the generalizations of Gleason's theorem.
Maeda is drawing on results by J. F. Aarnes \cite{Aar69,Aar70}, J. Gunson
\cite{Gun72}, E. Christensen \cite{Chr82,Chr85}, F. J. Yeadon
\cite{Yea83,Yea84}\ and K. Saito \cite{Sai85}. The proofs given in Maeda's
paper are by no means trivial. The central point of course is to show that a
quasi-state is linear on $\mathcal{R}_{sa}$ for \emph{non-commuting}
self-adjoint operators $A,B$. Maeda uses Gleason's theorem \cite{Gle57} for
the type $I_{n}$ algebras. Types $II$ and $III$ require a lot more work. We
cite the main result (Thm. 12.1 in \cite{Mae89}):

\begin{theorem}
\label{TGenGleas}(Christensen, Yeadon, Maeda et. al.) Let $\mathcal{R}$ be a
von Neumann algebra without direct summand of type $I_{2}$, and let $\mu$ be a
finitely additive probability measure on the complete orthomodular lattice
$\mathcal{P(R)}$. $\mu$ can be extended to a state $\widehat{\mu}$ of
$\mathcal{R}$, and moreover%
\[
\forall E,F\in\mathcal{P(R)}:|\mu(E)-\mu(F)|\leq||E-F||.
\]

\end{theorem}

It follows that the quasi-state $v^{\prime}$ is a state of $\mathcal{R}$ if
the von Neumann algebra $\mathcal{R}$ has no summand of type $I_{2}$. However,
the state $v^{\prime}$ is not normal necessarily, i.e. it need not be of the
form $v^{\prime}(\_)=tr(\rho\_)$, so we cannot use the same argument as
before. Instead, we will show that $v^{\prime}$ is a \emph{multiplicative}
state, using a result by J. Hamhalter (\cite{Ham93}), and give a second proof
of the Kochen-Specker theorem, valid for all von Neumann algebras without
summands of types $I_{1}$ and $I_{2}$.\newline

We will exploit the fact that $v^{\prime}|_{\mathcal{P(R)}}$ is a
\emph{two-valued} measure, i.e. $v^{\prime}(E)\in\{0,1\}$ for all
$E\in\mathcal{P(R)}$. From now on, measure will always mean finitely additive
probability measure. We cite Lemma 5.1 of \cite{Ham93} with proof:

\begin{lemma}
\label{LTwoValMeasMultState}Let $\mathcal{R}$ be a von Neumann algebra without
type $I_{2}$ summand. Every two-valued measure on $\mathcal{P(R)}$ can be
extended to a multiplicative state of $\mathcal{R}$.
\end{lemma}

\begin{proof}
Let $\mu$ be a two-valued measure on $\mathcal{P(R)}$. Using the
Gleason-Christensen-Yeadon theorem (Thm. \ref{TGenGleas}), we can extend $\mu$
to a state $\phi$ of $\mathcal{R}$. Let $\pi_{\phi}:\mathcal{R}\rightarrow
\mathcal{H}_{\phi}$ be the GNS representation engendered by $\phi$. Let
$x_{\phi}$ be a unit cyclic vector of $\pi_{\phi}$ such that $\phi
=\omega_{x_{\phi}}\circ\pi_{\phi}$, where $\omega_{x_{\phi}}$ is the vector
state given by $x_{\phi}$. For every $E\in\mathcal{P(R)}$ we have
$\mu(E)=\left\langle \pi_{\phi}(E)x_{\phi},x_{\phi}\right\rangle $. We see
that $\mu(E)$ is either $0$ or $1$. It follows that either $\pi_{\phi
}(E)x_{\phi}=x_{\phi}$ or $\pi_{\phi}(E)x_{\phi}=0$. Hence, $\mathcal{H}%
_{\phi}=\overline{\operatorname*{lin}}\{\pi_{\phi}(A)x_{\phi}\ |\ A\in
\mathcal{R}\}=\overline{\operatorname*{lin}}\{\pi_{\phi}(E)\ |\ E\in
\mathcal{P(R)}\}=\operatorname*{lin}\{x_{\phi}\}$, where $\operatorname*{lin}$
means the linear span and $\overline{\operatorname*{lin}}$ its closure.
Therefore, for every $A\in\mathcal{R}$ there is a complex number $\lambda_{A}$
such that $\pi_{\phi}(A)x_{\phi}=\lambda_{A}x_{\phi}$. Obviously,
$\lambda_{AB}=\lambda_{A}\lambda_{B}$ for $A,B\in\mathcal{R}$, and therefore%
\begin{align*}
\phi(AB)  &  =\left\langle \pi_{\phi}(AB)x_{\phi},x_{\phi}\right\rangle
=\lambda_{AB}\\
&  =\lambda_{A}\lambda_{B}=\left\langle \pi_{\phi}(A)x_{\phi},x_{\phi
}\right\rangle \left\langle \pi_{\phi}(B)x_{\phi},x_{\phi}\right\rangle \\
&  =\phi(A)\phi(B)
\end{align*}
for all $A,B\in\mathcal{R}$.
\end{proof}

\begin{corollary}
\label{CValStateMult}Let $\mathcal{R}$ be a von Neumann algebra without type
$I_{2}$ summand. The state $v^{\prime}$ induced by a valuation function
$v:\mathcal{R}_{sa}\rightarrow\mathbb{R}$ is multiplicative.
\end{corollary}

We now give a short proof (in two lemmata) for the well-known fact that a von
Neumann algebra $\mathcal{R}$ of fixed type has no multiplicative states
unless $\mathcal{R}$ is of type $I_{1}$, i.e. abelian. See also Thm. 5.3 in
\cite{Ham93}.

\begin{lemma}
\label{LTypeInNoMultStates}Let $\mathcal{R}$ be a von Neumann algebra of type
$I_{n},\ n\geq2$. There are no multiplicative states of $\mathcal{R}$.
\end{lemma}

\begin{proof}
If $\mathcal{R}$ is of type $I_{n}$, then $I$ is the sum of $n$ equivalent
abelian orthogonal projections $E_{j}\ (j=1,...,n)$. $E_{1}\sim E_{2}$ means
that there is a partial isometry $\theta\in\mathcal{R}$ such that
$E_{1}=\theta^{\ast}\theta$ and $E_{2}=\theta\theta^{\ast}$. Let $\phi$ be a
multiplicative state of $\mathcal{R}$. In particular, $\phi$ is a tracial
state, i.e. $\phi(AB)=\phi(BA)$ for all $A,B\in\mathcal{R}$, hence%
\[
\phi(E_{1})=\phi(\theta^{\ast}\theta)=\phi(\theta\theta^{\ast})=\phi(E_{2}).
\]
In the same manner, one obtains $\phi(E_{1})=\phi(E_{2})=\phi(E_{3}%
)=...=\phi(E_{n})$. But $\phi(E_{1})\in\{0,1\}$, since $\phi$ is
multiplicative, so%
\[
\phi(I)=\phi(\sum_{j=1}^{n}E_{j})=\sum_{j=1}^{n}\phi(E_{n})\in\{0,n\},
\]
which is a contradiction.
\end{proof}

\begin{lemma}
\label{LOtherTypesNoMultStates}Let $\mathcal{R}$ be a von Neumann algebra of
type $I_{\infty,}\ II$ or $III$. There are no multiplicative states of
$\mathcal{R}$.
\end{lemma}

\begin{proof}
First regard the case that $\mathcal{R}$ is of type $I_{\infty}$. Since
$\mathcal{R}$ is properly infinite, we can use the halving lemma (Lemma 6.3.3
in \cite{KadRinII97}) to show that there is a projection $F\in\mathcal{P(R)}$
such that $F\sim F^{\perp}:=I-F$. For $\mathcal{R}$ a type $II$ or $III$
algebra, we use lemma \ref{LPartingProjs} (choose $E=I$ and $n=2)$ to the same
effect. $F\sim F^{\perp}$ means that there is a partial isometry $\theta
\in\mathcal{R}$ such that $F=\theta^{\ast}\theta$ and $F^{\perp}=\theta
\theta^{\ast}$. Let $\phi$ be a multiplicative state of $\mathcal{R}$, so%
\[
\phi(F)=\phi(\theta^{\ast}\theta)=\phi(\theta\theta^{\ast})=\phi(F^{\perp}).
\]
Since $\phi(F)\in\{0,1\}$, we have%
\[
\phi(I)=\phi(I-F+F)=\phi(F^{\perp})+\phi(F)\in\{0,2\},
\]
which is a contradiction.
\end{proof}

\ 

Now let $\mathcal{R}$ be an arbitrary von Neumann algebra without summand of
type $I_{2}$. Let $P_{I_{1}}\in\mathcal{P(R)}$ be the maximal abelian central
projection, $P_{I}$ the maximal central projection such that $\mathcal{R}%
P_{I}$ is of type $I$, but has no central abelian portion, $P_{II}$ the
maximal central projection such that $\mathcal{R}P_{II}$ is of type $II$ and
$P_{III}$ the maximal central projection such that $\mathcal{R}P_{III}$ is of
type $III$. We have $I=P_{I_{1}}+P_{I}+P_{II}+P_{III}$ (see Thm. 6.5.2 in
\cite{KadRinII97}).\newline

Every projection $E\in\mathcal{P(R)}$ can be written as $E=E_{I_{1}}%
+E_{I}+E_{II}+E_{III}$ for orthogonal projections $E_{I_{1}}\in\mathcal{R}%
P_{I_{1}},\ E_{I}\in\mathcal{R}P_{1},\ E_{II}\in\mathcal{R}P_{II}$ and
$E_{III}\in\mathcal{R}P_{III}$. Let $v:\mathcal{R}_{sa}\rightarrow\mathbb{R}$
be a valuation function and let $v^{\prime}$ be the induced state of
$\mathcal{R}$. Since $v^{\prime}|_{\mathcal{P(R)}}$ is finitely additive,
$v^{\prime}|_{\mathcal{R}_{sa}}=v$ and $v(I)=1=v(P_{I_{1}})+v(P_{I}%
)+v(P_{II})+v(P_{III})$, exactly one term on the right hand side equals $1$,
the others are zero. Let $P_{x}\ (x\in\{I_{1},I,II,III\})$ denote the central
projection such that $v(P_{x})=1$. It follows that $v(E)=0$ for all $E\leq
P_{y}\ (y\in\{I_{1},I,II,III\})$ for all $y\neq x$, since $v^{\prime
}|_{\{E,P_{y}\}^{\prime\prime}}$ is positive. This means that the valuation
function is concentrated at $\mathcal{R}P_{x}$ in the sense that $v(E)=0$ for
all projections $E$ orthogonal to $P_{x}$.\newline

$v|_{\mathcal{R}P_{I}}$ cannot be a valuation function for $(\mathcal{R}%
P_{I})_{sa}$, since the induced state $(v|_{\mathcal{R}P_{I}})^{\prime}$ on
$\mathcal{R}P_{I}$ would be multiplicative, but $\mathcal{R}P_{I}$ is a sum of
type $I_{n}$ algebras, $n\in\{3,4,...,\infty\}$, and none of these algebras
has a multiplicative state. Similarly, $v|_{\mathcal{R}P_{II}}$ cannot be a
valuation function for $\mathcal{R}P_{II}$ and $v|_{\mathcal{R}P_{III}}$
cannot be a valuation function for $\mathcal{R}P_{III}$, because
$\mathcal{R}P_{II}$ and $\mathcal{R}P_{III}$ have no multiplicative states. It
follows that $v|_{\mathcal{R}P_{I_{1}}}$ is a valuation function for
$(\mathcal{R}P_{I_{1}})_{sa}$ and the induced multiplicative state
$(v|_{\mathcal{R}P_{I_{1}}})^{\prime}$ equals $v^{\prime}$. For the abelian
part $\mathcal{R}P_{I_{1}}$, a \textquotedblright hidden state
space\textquotedblright\ is given by the Gelfand spectrum $\Omega
(\mathcal{R}P_{I_{1}})$, each element $\omega\in\Omega(\mathcal{R}P_{I_{1}})$
is a hidden pure state and induces a valuation function, assigning a spectral
value to each $A\in\mathcal{R}P_{I_{1}}$ by evaluating $\omega(A)$, preserving
functional relations. We have shown that only in this trivial situation one
can have a valuation function. We obtain:

\begin{lemma}
\label{LNoValFuncGeneral}Let $\mathcal{R}$ be a von Neumann algebra without
type $I_{2}$ summand, and let $P_{I_{1}}\in\mathcal{P(R)}$ be the maximal
abelian central projection. There exists a valuation function $v:\mathcal{R}%
_{sa}\rightarrow\mathbb{R}$ if and only if $\mathcal{R}$ has a summand of type
$I_{1}$, i.e. $P_{I_{1}}\neq0$. In this case, $v=v|_{\mathcal{R}P_{I_{1}}}$,
and the valuation function $v$ is completely trivial on the non-abelian part
$\mathcal{R}(I-P_{I_{1}})$ of $\mathcal{R}$, $v|_{(\mathcal{R}(I-P_{I_{1}%
}))_{sa}}=0$.
\end{lemma}

Summing up, we have a generalized Kochen-Specker theorem:

\begin{theorem}
\label{TGenKS}Let $\mathcal{R}$ be a von Neumann algebra without type $I_{2}$
summand. If $\mathcal{R}$ has no type $I_{1}$ summand, then the Kochen-Specker
theorem holds. If $\mathcal{R}$ has a type $I_{1}$ summand, then there is a
hidden state space in the sense described in the introduction, but only for
the trivial, abelian part $\mathcal{R}P_{I_{1}}$ of $\mathcal{R}$.
\end{theorem}

\section{The presheaf perspective}

In a remarkable series of papers, C. J. Isham and J. Butterfield (with J.
Hamilton as co-author of the third paper) have given several reformulations of
the Kochen-Specker theorem (\cite{IshBut98,IshBut99,HIB00,IshBut02}). They use
the language of \emph{presheafs on a category}:

\begin{definition}
Let $\mathcal{C}$ be a small category. A \textbf{presheaf on }$\mathcal{C}$ is
a covariant functor%
\[
P:\mathcal{C}^{op}\longrightarrow\operatorname*{Set}\text{.}%
\]

\end{definition}

The observation is that the FUNC principle, condition (b) in Def.
\ref{DValFunc}, means that a certain square diagram commutes:

\begin{center}%
\setsqparms[1`1`1`1;1300`700]
\square[A`B`v(A)`v(B);g`v`v`g]%

\end{center}

This diagram captures the situation $B=g(A)$ and $v(B)=v(g(A))=g(v(A))$. Isham
and Butterfield observe that such a diagram can be read as expressing that
there is a \emph{section} of a presheaf on a category:

\begin{definition}
Let $P$ be a presheaf on a small category $\mathcal{C}$. A \textbf{global
section} $s$ of $P$ is a mapping $\mathcal{C}\rightarrow\operatorname*{Set}$
such that $s(a)\in P(a)$ for all $a\in\mathcal{C}$ and, whenever there is a
morphism $\varphi:a\rightarrow b$\ $(a,b\in\mathcal{C})$, the following
diagram commutes:%

\begin{center}
\setsqparms[1`1`1`-1;1300`700]
\square[a`b`s(a)`s(b);\varphi`s`s`P(\varphi)]
\end{center}%

\end{definition}

Please notice that the the horizontal arrow at the bottom is reversed, because
we are dealing with presheafs, i.e. contravariant functors $\mathcal{C}%
\rightarrow\operatorname*{Set}$.\newline

There are several choices for the category and the presheaf that can be used
to reformulate the Kochen-Specker theorem. We will generalize the proposal
made in \cite{HIB00}: Let $\mathfrak{A}\mathcal{(R)}$ denote the category of
unital abelian subalgebras of $\mathcal{R}$ (the unit of $\mathcal{M}%
\in\mathfrak{A}(\mathcal{R})$ is the unit of $\mathcal{R}$). A morphism
$\iota_{\mathcal{MN}}:\mathcal{M}\rightarrow\mathcal{N}$ exists whenever
$\mathcal{M\subseteq N}$. Hamilton, Isham and Butterfield only regard the case
$\mathcal{R=L(H)}$ and denote this category by $\mathcal{V}$.

\begin{definition}
(compare Def 2.3 in \cite{HIB00}) The \textbf{spectral presheaf} over
$\mathfrak{A}\mathcal{(R)}$ is the contravariant functor $\Sigma
:\mathfrak{A}\mathcal{(R)}\rightarrow\operatorname*{Set}$ defined as follows:

(i) On objects: $\Sigma(\mathcal{M}):=\Omega(\mathcal{M})$, the Gelfand
spectrum of $\mathcal{M}$.

(ii) On morphisms: If $\iota_{\mathcal{MN}}:\mathcal{M}\rightarrow\mathcal{N}$
is the inclusion, then $\Sigma(\iota_{\mathcal{MN}}):\Omega(\mathcal{N}%
)\rightarrow\Omega(\mathcal{M})$ is defined by $\Sigma(\iota_{\mathcal{MN}%
})(\omega):=\omega|_{\mathcal{M}}$.
\end{definition}

If there was a global section $s$ of $\Sigma$, the following diagram would commute:%

\begin{center}
\setsqparms[1`1`1`-1;1300`700]
\square[\mathcal{M}`\mathcal{N}`\Omega(\mathcal{M})`\Omega(\mathcal{N}%
);\iota_{\mathcal{MN}}`s`s`\Sigma(\iota_{\mathcal{MN}})]
\end{center}%
For $\mathcal{M}\in\mathfrak{A}(\mathcal{R})$, $s(\mathcal{M})\in
\Omega(\mathcal{M})$ and $s(\mathcal{M})=\Sigma(\iota_{\mathcal{MN}}%
)(s(\iota_{\mathcal{MN}}(\mathcal{M})),$ where $\iota_{\mathcal{MN}%
}(\mathcal{M})$ is the algebra $\mathcal{M}$ seen as part of $\mathcal{N}$ and
$s(\iota_{\mathcal{MN}}(\mathcal{M}))\in\Omega(\mathcal{N})$. The
commutativity of the diagram means that $s(\mathcal{M})$ is given as the
restriction of $s(\iota_{\mathcal{MN}}(\mathcal{M}))$ to $\Omega
(\mathcal{M})\subseteq\Omega(\mathcal{N})$.\newline

Such a choice of one element $s(\mathcal{M})$ of the Gelfand spectrum
$\Omega(\mathcal{M})$ per abelian subalgebra $\mathcal{M}$ of $\mathcal{R}$,
compatible with the spectral presheaf mappings, i.e. with restrictions
$\Omega(\mathcal{N})\rightarrow\Omega(\mathcal{M})$, would give a valuation
function when restricted to the self-adjoint elements: for all $A\in
\mathcal{M}_{sa}$, $s(\mathcal{M})(A)\in\operatorname*{sp}A$ and
$s(\mathcal{M})(f(A))=f(s(\mathcal{M})(A))$. \ The generalized Kochen-Specker
theorem (Thm. \ref{TGenKS}) hence shows that for von Neumann algebras
$\mathcal{R}$ without summands of types $I_{1}$ and $I_{2}$, there is no
global section of $\Sigma$.\newline

In Lemma \ref{LValFuncExtChar}, we saw that having a valuation function
$v:\mathcal{R}_{sa}\rightarrow\mathbb{R}$ would mean having a character
$v^{\prime}|_{\mathcal{M}}$ (an element of the Gelfand spectrum) for each
abelian subalgebra $\mathcal{M}\in\mathfrak{A}(\mathcal{R})$. It follows from
the FUNC principle that these characters are subject to the same conditions as
above: if $\mathcal{M}\subseteq\mathcal{N}$, then restricting $v^{\prime
}|_{\mathcal{N}}$ to $\Omega(\mathcal{M})$ must give $v^{\prime}%
|_{\mathcal{M}}$ (which is not possible globally). This choice of a category
and a presheaf thus brings the presheaf formulation of the Kochen-Specker
theorem very close to our first proof.\newline

There is a closely related formulation, using Stone spectra instead of Gelfand
spectra (see Def. \ref{DBoolSecStoneSpec}):

\begin{definition}
The \textbf{state presheaf} $\mathcal{M}^{1}$ on $\mathfrak{A}\mathcal{(R)}$
is defined as follows:

(i) On objects: $\mathcal{M}^{1}(\mathcal{M}):=\mathcal{M}^{1}(\mathcal{Q(M))}%
$, the set of positive Radon measures of norm $1$ on $\mathcal{Q(M)}$.

(ii) On morphisms: for $\mathcal{M},\mathcal{N}\in\mathfrak{A}\mathcal{(R)}$
such that $\mathcal{M}\subseteq\mathcal{N}$ let%
\begin{align*}
p_{\mathcal{M}}^{\mathcal{N}}:\mathcal{M}^{1}(\mathcal{N)}  &  \longrightarrow
\mathcal{M}^{1}(\mathcal{M)}\\
\mu_{\mathcal{N}}  &  \longmapsto p_{\mathcal{M}}^{\mathcal{N}}.\mu
_{\mathcal{N}},
\end{align*}
where $p_{\mathcal{M}}^{\mathcal{N}}.\mu_{\mathcal{N}}$ is the image measure
defined by%
\[
(p_{\mathcal{M}}^{\mathcal{N}}.\mu_{\mathcal{N}})(U):=\mu_{\mathcal{N}%
}((p_{\mathcal{M}}^{\mathcal{N}})^{-1}(U)),
\]
where $U\subseteq\mathcal{Q(M)}$ is a Borel set and $p_{\mathcal{M}%
}^{\mathcal{N}}:\mathcal{Q(}\mathcal{N}\mathcal{)}\rightarrow\mathcal{Q(}%
\mathcal{M}\mathcal{)},\ \beta\mapsto\beta\cap\mathcal{M}$ is the restriction
map between the Stone spectra.
\end{definition}

$\mathcal{M}^{1}$ really is a presheaf on $\mathfrak{A}\mathcal{(R)}$, since
obviously $p_{\mathcal{M}}^{\mathcal{M}}=id_{\mathcal{M}^{1}(\mathcal{M)}}$
and since $p_{\mathcal{M}}^{\mathcal{C}}=p_{\mathcal{M}}^{\mathcal{N}}\circ
p_{\mathcal{N}}^{\mathcal{C}}$ as mappings $\mathcal{Q(P)\rightarrow
Q(N)\rightarrow Q(M)}$, the same holds for the mappings $\mathcal{M}%
^{1}(\mathcal{P)\rightarrow M}^{1}(\mathcal{N}\mathcal{)\rightarrow M}%
^{1}(\mathcal{M}\mathcal{)}$. Let $\mathcal{R}$ have no type $I_{1}$ and
$I_{2}$ summands. From the generalized Kochen-Specker theorem (Thm.
\ref{TGenKS}) it follows that$\ \mathcal{M}^{1}$ has no global sections
consisting entirely of point measures. The fact that $\mathcal{M}^{1}$ has no
such global sections is equivalent to the generalized Kochen-Specker theorem,
since a valuation function would induce a quasi-state $v^{\prime}$ (Lemma
\ref{LValFuncExtQS}) such that $v^{\prime}|_{\mathcal{M}}$ is a pure state for
every $\mathcal{M}\in\mathfrak{A}\mathcal{(R)}$ (and hence gives a point
measure on $\mathcal{Q(M)}$).\newline

This presheaf formulation emphasizes the fact that a valuation function would
give a \emph{pure} state of every $\mathcal{M}\in\mathfrak{A}(\mathcal{R})$.
The presheaf $\mathcal{M}^{1}$ does have global sections (every state of
$\mathcal{R}$ induces one, obviously), but it has no global sections
consisting entirely of point measures.

\section{Discussion}

We have presented two functional analytic proofs for the fact that there are
no valuation functions $v:\mathcal{R}_{sa}\rightarrow\mathbb{R}$ for
$\mathcal{R}$ a von Neumann algebra. The first proof only uses Gleason's
classical theorem (Thm. \ref{TGleason}) and holds for $\mathcal{R}$ a type
$I_{n}$ factor, $n\geq3$. The second proof depends on the
Gleason-Christensen-Yeadon theorem (Thm. \ref{TGenGleas}) and holds for von
Neumann algebras $\mathcal{R}$ without summands of types $I_{1}$ and $I_{2}$.
To the best of our knowledge, for the first time von Neumann algebras other
than the type $I_{n}$ factors $\mathcal{L(H)}$ have been treated. The
generalized Kochen-Specker theorem follows: there is no hidden states model of
quantum theory in the sense described in the introduction.\newline

Both proofs are based on the fact that having a valuation function $v$ would
mean having a \emph{state} $v^{\prime}$ of $\mathcal{R}$, which follows from
Gleason's theorem, and this state has properties that lead to a contradiction.
In the first proof, for type $I_{n}$ factors ($n\geq3$), the state is of the
form $v^{\prime}(\_)=tr(\rho\_)$, so there are projections $E\in
\mathcal{P(R)}$ such that $v^{\prime}(E)\notin\{0,1\}$. The second proof,
which is much more general, uses the fact that $v^{\prime}$ is a
multiplicative state. Since there are no multiplicative states except on type
$I_{1}$, i.e. abelian, algebras, the Kochen-Specker theorem holds for all von
Neumann algebras without summands of types $I_{1}$ and $I_{2}$. Type $I_{2}$
must be excluded since Gleason's theorem and the Gleason-Christensen-Yeadon
theorem only hold if $\mathcal{R}$ has no type $I_{2}$ summand. It is known
that every type $I_{2}$ algebra admits a two-valued measure and hence a
valuation function, see Rem. 5.4 in \cite{Ham93}. If $\mathcal{R}$ has a type
$I_{1}$ summand, then there are valuation functions, but they are concentrated
at the trivial, abelian part $\mathcal{R}P_{I_{1}}$ of $\mathcal{R}$, where
$P_{I_{1}}\in\mathcal{P(R)}$ is the maximal abelian central
projection.\newline

The fact that the defining conditions of a valuation function $v$ inevitably
lead to a multiplicative state shows that these conditions are very strong.
Indeed, although the FUNC principle only seems to pose conditions on commuting
operators, this is not the case: $v^{\prime}$ is a state, i.e. additive on
non-commuting operators also. In the physics literature, an abelian subalgebra
$\mathcal{M}$ of $\mathcal{R}$ is called a \emph{context}. Of course, the
contexts give nothing like a partition of $\mathcal{R}$ into abelian,
\textquotedblright classical\textquotedblright\ parts, but are interwoven in
an intricate manner, since an observable $A\in\mathcal{R}_{sa}$ typically is
contained in many abelian subalgebras. The FUNC principle poses conditions
within each context, but since typically $A=f(B)=g(C)$ for non-commuting
observables $B,C$, it also poses conditions on non-commuting observables,
across contexts. The presheaf formulations presented in section 3 clearly show
that the Kochen-Specker theorem means that there is no state $\phi$ of
$\mathcal{R}$ such that for all contexts $\mathcal{M}\in\mathfrak{A}%
(\mathcal{R})$, the restriction $\phi|_{\mathcal{M}}$ is a pure state.\newline

The Kochen-Specker theorem is little more than a corollary to Gleason's
theorem, in a more general sense than worked out by Bell (\cite{Bell66}). The
fact that $v^{\prime}$ is a state comes from Gleason's theorem (or its
generalization): a valuation function $v$ defines a quasi-state $v^{\prime}$
in a canonical manner, and restricting the quasi-state $v^{\prime}$ to the
projection lattice $\mathcal{P(R)}$ gives a finitely additive probability
measure. Gleason's theorem shows that $v^{\prime}$ must be a state. The deep
meaning of Gleason's theorem is that the simple, \emph{lattice-theoretic}
condition of finite additivity on each distributive sublattice, which is a
condition on finite joins actually ($E+F=E\vee F$ for orthogonal projections
$E,F$), suffices to guarantee additivity of the functional $v^{\prime}$
defined by linear extension of the probability measure (and taking the
appropriate limit, see e.g. Ch. III.7 of \cite{Mae89}). Of course, finite
additivity on each distributive sublattice is a condition across distributive
sublattices, since each projection $E$ is contained in many distributive
sublattices.\newline

But the defining conditions of a valuation function $v$ are even stronger:
using the fact that $v(E)\in\operatorname*{sp}E=\{0,1\}\ (E\in\mathcal{P(R)}%
)$, we saw that the state $v^{\prime}$ is multiplicative, which is only
possible if $v$ is concentrated at the abelian part $\mathcal{R}P_{I_{1}}$ of
$\mathcal{R}$. Thus, a valuation function and a hidden states model can only
exist for the trivial, abelian situation. This generalizes to arbitrary
von\ Neumann algebras a result found by J. D. Malley (\cite{Mal04}). It also
rebuts the critique of von Neumann's proof from 1932 (\cite{vNeu32}). Von
Neumann posed additivity conditions on non-commuting observables, which was
strongly criticized by Bell (\cite{Bell66}) as unphysical. Of course, the
Gelfand representation of an abelian von Neumann algebra is a hidden states
model, the Gelfand spectrum $\Omega(\mathcal{R})$ taking the r\^{o}le of the
\textquotedblright hidden\textquotedblright\ state space.\newline

We have shown more than the fact that there are no non-trivial hidden states
models: Each element $\omega$ of a hidden state space $\Omega$ would give a
valuation function, as described in the introduction, but having a valuation
function would not necessarily mean having a hidden states model. A valuation
function would simply assign values to all observables in a manner consistent
with the FUNC principle, which would be an important piece of a realist
quantum theory. Since we have ruled out this possibility, there are no such
na\"{\i}ve realist models of quantum theory.

\subsection{Acknowledgements}

The author gratefully acknowledges Prof. H. de Grootes constant interest in
his work, numerous discussions and continuous support. The author was
supported by the Studienstiftung des Deutschen Volkes.

\end{document}